LA ROCCA PATRICK[1], BONASIA MATTEO[1], MOREO PRINCE[1] AND ZAMARIOLA CRISTINA[1]
BENNA CARLO[2], GARDIOL DANIELE[2] AND PETTITI GIUSEPPE[2]

1) IIS Curie Vittorini, Corso Allamano 130, 10095, Grugliasco (TO), Italy, TOIS03400P@istruzione.it

2) INAF-Osservatorio Astrofisico di Torino, via Osservatorio 20, I-10025 Pino Torinese (TO), Italy, giuseppe.pettiti@inaf.it

**Abstract.** This study aims to assess the properties and classification of 62 variable stars in Cygnus, little studied since their discovery and originally reported in the Information Bulletin on Variable Stars (IBVS) 1302.

Using data from previous studies and several astronomical databases, we performed our analysis mainly utilizing a period analysis software and comparing the photometric characteristics of the variables in a Colour-Absolute Magnitude Diagram.

For all stars, the variability is confirmed. We discovered new significant results for the period and/or type of 17 variables and highlighted incorrect cross-reference names on astronomical databases for 23 stars. For 3 stars, whose original type was unknown, we propose a new type. We calculated an epoch of a minimum or a maximum for 24 stars; for 3 of them, the epoch has been defined for the first time.

This assessment also identifies some cases for which results from the ASAS-SN Catalog of Variable Stars are not consistent with results from Gaia DR2 and/or our analysis.

## 1. Introduction

The IBVS 1302 (Maffei, 1977) contains the results of a systematic photographic search of variable stars in a field centred on the star γ Cygni, performed at the Astrophysical Observatory of Asiago from 1967 to 1974. The survey identified 62 new variables whose provisional name, original coordinates, final designation, and main characteristics are shown in Table 1. They are mostly late spectral type stars, with low effective temperature and maximum luminosity at infrared wavelengths, that are currently classified as Mira, Semiregular or Irregular variable. The original coordinates of 3 of the 62 variables have been revised in the later IBVS 6242 (Nesci, 2018).

Initial classification and properties of variable stars discovered through surveys are sometimes affected by mistakes or uncertainties that remain undiscovered or unresolved due to the lack of new observations or further analysis.

The variable stars of IBVS 1302 are generally faint and little studied since their discovery.

In this study, we assess the original physical and photometric characteristics and the classification of the variables, based on data mined by more recent and public astronomical articles and databases.

In this paper, when the term 'original' is used for the characteristics of the variables, we refer to the data defined in IBVS 1302 (Maffei, 1977).

Section 2 describes the method of our assessment. In section 2.1, we provide a general assessment of the classification type of the variables based on their Gaia photometric characteristics; a more detailed analysis for each variable, based on the results of previous studies and of our photometric analysis, is provided in section 2.2. Our conclusions are summarized in section 3.

Table 1     Variable stars from IBVS 1302

| Provisional name | Original coordinates R.A. - Decl. (1950.0) | Variable Designation | Type | Mag. Range (I-N + RG5) | Period (d) |
|---|---|---|---|---|---|
| M 210 | 20h 27m 16.1s  +39° 32' 41" | V1659 Cyg | M | 12.1 - 15.4 | 770 |
| M 211 | 29 12.2  41 35 02 | V1660 Cyg | M | 13.0 - 15.7 | 420 |
| M 212 | 21 11.0  41 07 19 | V1692 Cyg | SR | 14.7 - 15.7 | 450: |
| M 213 | 19 41.2  41 23 00 | V1688 Cyg | M | 13.5 - 15.6 | 312: |
| M 214 | 19 52.7  41 27 24 | NSV 25117 | - | 14.8 -16.0 | - |
| M 215 | 21 11.0  41 41 23 | V1693 Cyg | SR or M | 14.1 - 16.0 | 400: |
| M 216 | 23 41.0  42 16 37 | V1658 Cyg | M | 11.1 - 14.5 | 382 |
| M 217 | 23 35.1  42 22 29 | V1656 Cyg | M | 11.5 - 14.1 | 308 |
| M 218 | 21 54.5  42 40 57 | V1694 Cyg | SR or M | 14.4 - 15.6 | 308:: |
| M 219 | 20 47.4  42 39 54 | V1690 Cyg | RV | 11.3 - 14.2 | 285 |
| M 220 | 18 20.5  42 53 23 | V1684 Cyg | SR | 13.0 - 14.2 | 400:: |
| M 221 | 16 08.0  42 22 19 | V1681 Cyg | SR | 14.7 - 15.9 | 280: |
| M 222 | 16 50.8  41 58 46 | V1651 Cyg | M | 13.4 - 15.6 | 383 |
| M 223 | 19 50.4  42 16 24 | V1689 Cyg | SR or M | 14.2 - 15.9 | 484: |
| M 224 | 17 23.9  41 37 24 | V1682 Cyg | SR | 14.8 - 15.7 | 290:: |
| M 225 | 13 51.5  42 12 43 | V1680 Cyg | SR | 15.0 - 15.8 | - |
| M 226 | 27 51.9  38 55 59 | NSV 25162 | - | 15.0 - 15.7 | - |
| M 228 | 23 33.3  38 32 44 | V1655 Cyg | M | 12.0 - 14.8 | 462 |
| M 229 | 23 45.6  38 09 17 | V1657 Cyg | M | 13.0 - 15.1 | 310 |
| M 230 | 21 13.4  38 54 34 | V1691 Cyg | M: | 13.3 - 15.4 | 385: |
| M 233 | 18 54.1  39 13 05 | V1653 Cyg | M | 10.4 - 12.2 | 335 |
| M 234 | 20 30.1  40 33 33 | V1654 Cyg | M | 10.1 - 14.4 | 268: |
| M 235 | 16 42.1  40 24 41 | V1649 Cyg | M | 12.5 - 15.0 | 442 |
| M 236 | 11 19.2  40 43 11 | V1638 Cyg | SR | 12.8 - 14.6 | 416 |
| M 237 | 11 07.1  40 54 01 | V1637 Cyg | SR | 12.7 - 14.5 | 410 |
| M 238 | 10 16.6  40 51 19 | V1632 Cyg | SR | 12.3 - 14.7 | 318 |
| M 239 | 09 07.0  41 06 57 | V1628 Cyg | M | 12.1 - 14.5 | 261 |
| M 240 | 08 46.6  41 13 47 | V1627 Cyg | M | 12.6 - 14.4 | 300 |
| M 241 | 10 25.6  41 13 19 | NSV 25055 | L:: | 12.4 - 13.5 | - |
| M 242 | 10 34.5  41 16 20 | V1677 Cyg | I | 12.5 - 13.4 | - |
| M 243 | 10 07.2  41 23 51 | V1631 Cyg | M | 11.5 - 13.6 | 430 |
| M 244 | 10 08.9  41 27 02 | V2311 Cyg | E | 12.5 - 13.4 | - |
| M 245 | 10 59.6  41 26 57 | V1636 Cyg | M | 10.7 - 14.2 | 322 |
| M 246 | 10 36.0  41 42 45 | V1634 Cyg | M | 11.9 - 13.7 | 251 |
| M 247 | 13 21.8  41 08 25 | NSV 25072 | C:: | 12.8 - 13.8 | - |
| M 248 | 14 49.4  41 25 48 | V1647 Cyg | SR or M | 13.6 - 14.8 | 334: |
| M 249 | 21 55.2  38 20 18 | V1322 Cyg | L:: | 7.2 - 8.6 | - |

| Provisional name | Original coordinates R.A. - Decl. (1950.0) | Variable Designation | Type | Mag. Range (I-N + RG5) | Period (d) |
|---|---|---|---|---|---|
| M 251 | 19 29.7  38 53 19 | NSV 25113 | - | 14.4 - 15.3 | - |
| M 252 | 16 49.7  39 08 22 | V1650 Cyg | M | 9.3 - 11.3 | 395 |
| M 253 | 16 09.6  38 57 36 | V1648 Cyg | M | 13.6 - 15.2 | 300 |
| M 254 | 17 15.1  38 45 10 | NSV 13006 | - | 14.4 - 15.6 | - |
| M 256 | 17 24.9  38 21 32 | V1652 Cyg | M | 13.2 - 15.2 | 565: |
| M 257 | 13 50.6  38 16 23 | V433 Cyg | M | 8.7 - 11.4 | 400 |
| M 258 | 12 31.1  38 18 35 | V1643 Cyg | M | 11.1 - 14.6 | 297 |
| M 259 | 13 32.9  39 49 32 | V1646 Cyg | M: | 12.4 - 15.5 | 388 |
| M 260 | 13 10.2  39 47 00 | NSV 25071 | E: | 13.2 - 14.8 | - |
| M 261 | 12 00.6  40 13 10 | V1642 Cyg | M | 13.1 - 14.8 | 290 |
| M 263 | 09 31.5  40 55 35 | V1629 Cyg | SR | 13.8 - 14.9 | 460: |
| M 264 | 09 33.9  40 43 28 | V1630 Cyg | M | 12.4 - 13.8 | 415 |
| M 265 | 07 41.0  40 34 53 | V1625 Cyg | SR: | 12.2 - 14.5 | - |
| M 266 | 05 47.6  40 34 59 | V1622 Cyg | M | 10.4 - 14.3 | 350 |
| M 267 | 05 51.0  39 47 49 | V1623 Cyg | M | 11.9 - 15.0 | 310 |
| M 268 | 07 55.0  39 37 18 | V1626 Cyg | SR:: | 12.9 - 15.1 | 400 |
| M 269 | 07 01.4  39 04 04 | NSV 25019 | L:: | 13.4 - 14.2 | - |
| M 270 | 10 20.6  39 27 47 | V1633 Cyg | M | 8.7 - 11.3 | 404 |
| M 271 | 11 36.1  39 17 52 | V1640 Cyg | M | 12.7 - 15.4 | 412 |
| M 272 | 12 50.6  39 04 01 | V1645 Cyg | M | 12.3 - 15.0 | 380 |
| M 273 | 10 46.4  38 53 27 | V1635 Cyg | M | 12.4 - 15.1 | 400 |
| M 274 | 11 59.7  38 29 00 | V1641 Cyg | SR or M | 13.1 - 15.4 | 584:: |
| M 275 | 11 34.4  38 29 39 | V1639 Cyg | M | 11.9 - 15.2 | 360 |
| M 276 | 11 37.7  38 25 24 | V1678 Cyg | E | 12.4 - 14.4 | - |
| M 277 | 17 36.5  40 46 43 | V1683 Cyg | I | 13.2 - 14.7 | - |

## 2. Data analysis

Searching by the provisional names assigned by Maffei, for each star we got the name defined in the variable stars designation system, as well as the J2000.0 equatorial coordinates, from the SIMBAD (Set of Identifications, Measurements, and Bibliography for Astronomical Data) astronomical database. This basic information allowed us to look for physical, spectroscopic, and photometric data available in other astronomical databases and catalogues, specifically the ones from the American Association of Variable Stars Observers (AAVSO, Kafka 2020), the All-Sky Automated Survey for Supernovae (ASAS-SN, Kochanek et al. 2017; Jayasinghe et al. 2018b) and the Gaia Data Release 2 (Gaia DR2, Gaia Collaboration, 2018 and Bailer-Jones et al. 2018). In Table 2 we list, for each variable, the Gaia DR2 source ID, the apparent median Gaia G magnitude, and the absolute magnitude $M_G$ with associated errors.

The absolute magnitudes were calculated using the distances reported in Table 3, which also summarises the effective temperature Teff and the Gaia colour index Bp-Rp, available from Gaia DR2. The interstellar extinction or circumstellar extinction was not considered in the absolute magnitude calculation because the extinction value estimation in the filter G is available only for 17 stars of IBVS 1302.

Table 2     Gaia DR2 source ID, apparent and absolute magnitudes

| Name | Gaia DR2 Source ID | Gaia DR2 Median G (mag.) | ΔG (±) | Absolute magnitude $M_G$ | Δ$M_G$ (-) | Δ$M_G$ (+) |
|---|---|---|---|---|---|---|
| NSV 13006 | 2061308294621233536 | 16.092 | - | 2.76 | 0.49 | 0.42 |
| NSV 25019 | 2061970956535372160 | 14.001 | - | 0.09 | 0.80 | 0.72 |
| NSV 25055 | 2074619978810093952 | 14.147 | 0.009 | 1.61 | 0.92 | 0.72 |
| NSV 25071 | 2062218896403414784 | 14.226 | - | 2.15 | 0.14 | 0.15 |
| NSV 25072 | 2062620870978142592 | 13.806 | - | 5.32 | 0.04 | 0.04 |
| NSV 25113 | 2061392957003016704 | 16.855 | - | 3.22 | 0.73 | 0.72 |
| NSV 25117 | 2068494908757889024 | 16.838 | - | 5.25 | 1.02 | 0.78 |
| NSV 25162 | 2064177229676836608 | 16.151 | 0.001 | 5.65 | 0.13 | 0.12 |
| V433 Cyg | 2060941710574890368 | 12.215 | 0.032 | 0.36 | 0.98 | 0.75 |
| V1322 Cyg | 2061130757837456000 | 8.782 | 0.001 | -2.04 | 0.08 | 0.08 |
| V1622 Cyg | 2074170102454813056 | 13.916 | 0.043 | -0.17 | 0.84 | 0.78 |
| V1623 Cyg | 2074087952635187840 | 14.324 | 0.046 | 2.17 | 0.77 | 0.59 |
| V1625 Cyg | 2074497898646252672 | 14.047 | - | 1.09 | 0.74 | 0.60 |
| V1626 Cyg | 2062063762193656832 | 14.861 | 0.029 | 1.55 | 0.94 | 0.81 |
| V1627 Cyg | 2074587371436708352 | 14.651 | 0.033 | 0.85 | 0.59 | 0.50 |
| V1628 Cyg | 2074538958561467008 | 15.330 | 0.054 | 1.06 | 0.76 | 0.74 |
| V1629 Cyg | 2074522259729145344 | 15.665 | 0.030 | 1.68 | 0.76 | 0.70 |
| V1630 Cyg | 2074515662659100800 | 14.389 | 0.027 | 1.68 | 0.62 | 0.51 |
| V1631 Cyg | 2074642209560870656 | 14.463 | 0.034 | 2.12 | 0.79 | 0.62 |
| V1632 Cyg | 2074517415005650944 | 15.307 | - | 3.79 | 1.70 | 1.29 |
| V1633 Cyg | 2062369288973870976 | 10.201 | 0.024 | -0.88 | 0.26 | 0.23 |
| V1634 Cyg | 2074717423030592512 | 14.570 | 0.030 | 0.38 | 0.79 | 0.81 |
| V1635 Cyg | 2061782665176034816 | 14.582 | 0.046 | 1.57 | 0.98 | 0.86 |
| V1636 Cyg | 2074640560281165184 | 14.624 | 0.063 | 0.80 | 0.84 | 0.82 |
| V1637 Cyg | 2062602557236419456 | 15.047 | 0.048 | 0.86 | 0.71 | 0.66 |
| V1638 Cyg | 2062596406843140608 | 15.814 | 0.038 | 2.04 | 0.84 | 0.81 |
| V1639 Cyg | 2061718996577182848 | 14.725 | 0.051 | 2.66 | 0.99 | 0.75 |
| V1640 Cyg | 2062176702643423232 | 15.237 | 0.054 | 2.35 | 0.99 | 0.86 |
| V1641 Cyg | 2061712193348895872 | 16.208 | 0.045 | - | - | - |
| V1642 Cyg | 2062431273943088896 | 15.544 | 0.042 | 1.55 | 0.77 | 0.74 |
| V1643 Cyg | 2060950884625235584 | 14.535 | 0.050 | 1.95 | 1.00 | 0.82 |
| V1645 Cyg | 2062116882336948736 | 14.906 | 0.027 | 1.97 | 0.91 | 0.79 |
| V1646 Cyg | 2062219927195606016 | 15.755 | 0.053 | 4.36 | 1.12 | 0.81 |
| V1647 Cyg | 2068633344148880384 | 18.199 | 0.037 | 6.80 | 0.89 | 0.66 |
| V1648 Cyg | 2061373749925363968 | 15.679 | 0.047 | 2.67 | 0.86 | 0.77 |
| V1649 Cyg | 2062342728892073088 | 15.840 | 0.046 | 5.06 | 0.82 | 0.61 |
| V1650 Cyg | 2061470128971060864 | 12.212 | 0.018 | 0.18 | 0.89 | 0.71 |
| V1651 Cyg | 2068676225101943552 | 14.975 | 0.054 | 3.45 | 0.98 | 0.73 |

| Name | Gaia DR2 Source ID | Gaia DR2 Median G (mag.) | ΔG (±) | Absolute magnitude $M_G$ | Δ$M_G$ (-) | Δ$M_G$ (+) |
|---|---|---|---|---|---|---|
| V1652 Cyg | 2061283074573426432 | 15.996 | 0.073 | 5.93 | 0.44 | 0.37 |
| V1653 Cyg | 2061416012387664512 | 12.843 | 0.018 | -0.28 | 0.74 | 0.80 |
| V1654 Cyg | 2068314073455293440 | 14.085 | 0.041 | 0.46 | 0.72 | 0.70 |
| V1655 Cyg | 2061149896212437504 | 14.923 | 0.046 | 2.79 | 1.01 | 0.94 |
| V1656 Cyg | 2068922721865414400 | 13.584 | 0.028 | 0.62 | 0.81 | 0.73 |
| V1657 Cyg | 2058131118315524224 | 15.182 | 0.038 | 3.51 | 0.83 | 0.66 |
| V1658 Cyg | 2068873587439520768 | 13.621 | 0.037 | 1.63 | 0.87 | 0.70 |
| V1659 Cyg | 2067300014496616704 | 14.475 | 0.033 | 1.86 | 0.82 | 0.79 |
| V1660 Cyg | 2068042253564762240 | 15.036 | 0.038 | 3.84 | 0.96 | 0.75 |
| V1677 Cyg | 2074625991764380928 | 13.912 | 0.007 | 0.16 | 0.86 | 0.82 |
| V1678 Cyg | 2061706455272623744 | 13.682 | 0.019 | 2.27 | 0.47 | 0.40 |
| V1680 Cyg | 2080703026884997888 | 17.037 | 0.049 | 4.70 | 1.12 | 0.92 |
| V1681 Cyg | 2068789642295333120 | 16.722 | 0.050 | 3.28 | 0.86 | 0.82 |
| V1682 Cyg | 2068603416816874880 | 15.761 | 0.036 | 3.71 | 0.90 | 0.71 |
| V1683 Cyg | 2068356129775425024 | 15.213 | 0.023 | 5.28 | 0.06 | 0.06 |
| V1684 Cyg | 2068820948319278976 | 13.539 | 0.035 | -0.44 | 0.64 | 0.68 |
| V1688 Cyg | 2068400110240743552 | 15.390 | 0.048 | 2.93 | 0.95 | 0.83 |
| V1689 Cyg | 2068742161438598144 | 15.318 | 0.022 | 1.63 | 0.77 | 0.76 |
| V1690 Cyg | 2069127913922504704 | 13.725 | - | 3.36 | 0.48 | 0.40 |
| V1691 Cyg | 2061227136905659392 | 16.023 | 0.045 | 3.59 | 0.89 | 0.78 |
| V1692 Cyg | 2068437046957954176 | 17.976 | 0.027 | 6.60 | 1.09 | 0.83 |
| V1693 Cyg | 2068502158662430208 | 16.054 | 0.042 | 4.00 | 1.04 | 0.88 |
| V1694 Cyg | 2068944162340771328 | 15.557 | 0.022 | 3.60 | 0.99 | 0.79 |
| V2311 Cyg | 2074642999834894592 | 13.318 | 0.005 | 2.84 | 0.06 | 0.06 |

Table 3     Gaia DR2 Bp-Rp colour index, effective temperature Teff and distance

| Name | Bp-Rp | Teff (K) | Δ Teff (-) | Δ Teff (+) | Distance d (pc) | Δd (pc) (-) | Δd (pc) (+) |
|---|---|---|---|---|---|---|---|
| NSV 13006 | 2.493 | 3431 | 143 | 941 | 4631 | 815 | 1165 |
| NSV 25019 | 4.259 | 3284 | 9 | 64 | 6049 | 1698 | 2706 |
| NSV 25055 | 5.521 | 3283 | 15 | 43 | 3214 | 909 | 1690 |
| NSV 25071 | 1.326 | 4737 | 158 | 217 | 2596 | 159 | 181 |
| NSV 25072 | 0.978 | 5332 | 80 | 18 | 498 | 8 | 9 |
| NSV 25113 | 5.042 | 3326 | 49 | 92 | 5328 | 1503 | 2138 |
| NSV 25117 | 5.985 | 3284 | 8 | 39 | 2080 | 630 | 1251 |
| NSV 25162 | 1.887 | 3854 | 88 | 490 | 1259 | 69 | 77 |
| V433 Cyg | 5.710 | 3284 | 7 | 51 | 2351 | 684 | 1340 |
| V1322 Cyg | 1.371 | 4828 | 143 | 136 | 1461 | 51 | 55 |

| Name | Bp-Rp | Teff (K) | Δ Teff (-) | Δ Teff (+) | Distance d (pc) | Δd (pc) (-) | Δd (pc) (+) |
|---|---|---|---|---|---|---|---|
| V1622 Cyg | 5.394 | 3286 | 10 | 82 | 6578 | 1988 | 3087 |
| V1623 Cyg | 5.332 | 3293 | 16 | 84 | 2694 | 645 | 1150 |
| V1625 Cyg | 3.865 | 3335 | 42 | 670 | 3907 | 944 | 1599 |
| V1626 Cyg | 4.822 | 3301 | 8 | 102 | 4599 | 1435 | 2480 |
| V1627 Cyg | 3.285 | 4202 | 419 | 487 | 5759 | 1185 | 1795 |
| V1628 Cyg | 4.870 | 3316 | 41 | 96 | 7145 | 2060 | 3013 |
| V1629 Cyg | 4.230 | 3284 | 6 | 14 | 6278 | 1721 | 2639 |
| V1630 Cyg | 4.157 | 3285 | 5 | 72 | 3485 | 726 | 1159 |
| V1631 Cyg | 4.376 | 3291 | 15 | 7 | 2939 | 730 | 1295 |
| V1632 Cyg | 3.703 | 3690 | 242 | 587 | 2014 | 902 | 2383 |
| V1633 Cyg | 3.635 | 3295 | 9 | 126 | 1646 | 166 | 207 |
| V1634 Cyg | 2.594 | 4465 | 298 | 409 | 6895 | 2143 | 3043 |
| V1635 Cyg | 5.684 | 3279 | 11 | 37 | 4008 | 1312 | 2291 |
| V1636 Cyg | 6.433 | 3280 | 4 | 8 | 5821 | 1828 | 2758 |
| V1637 Cyg | 4.143 | 3286 | 9 | 104 | 6879 | 1806 | 2652 |
| V1638 Cyg | 6.047 | 3284 | 10 | 48 | 5678 | 1764 | 2668 |
| V1639 Cyg | 6.266 | 3284 | 9 | 48 | 2591 | 760 | 1492 |
| V1640 Cyg | 6.091 | 3284 | 9 | 52 | 3778 | 1240 | 2187 |
| V1641 Cyg | 5.999 | 3281 | 6 | 9 | - | - | - |
| V1642 Cyg | 5.094 | 3285 | 18 | 92 | 6297 | 1823 | 2675 |
| V1643 Cyg | 6.233 | 3282 | 6 | 11 | 3282 | 1037 | 1913 |
| V1645 Cyg | 6.175 | 3283 | 11 | 41 | 3873 | 1180 | 2026 |
| V1646 Cyg | - | - | - | - | 1905 | 594 | 1293 |
| V1647 Cyg | 3.797 | - | - | - | 1908 | 502 | 962 |
| V1648 Cyg | 5.199 | 3323 | 44 | 83 | 4002 | 1191 | 1933 |
| V1649 Cyg | 6.615 | 3280 | 5 | 7 | 1431 | 353 | 658 |
| V1650 Cyg | 6.128 | 3281 | 5 | 8 | 2547 | 708 | 1294 |
| V1651 Cyg | 6.663 | 3282 | 6 | 37 | 2015 | 574 | 1143 |
| V1652 Cyg | 6.063 | 3281 | 6 | 8 | 1031 | 162 | 234 |
| V1653 Cyg | 6.420 | 3281 | 5 | 8 | 4205 | 1212 | 1872 |
| V1654 Cyg | 6.010 | 3281 | 6 | 6 | 5304 | 1465 | 2091 |
| V1655 Cyg | 7.383 | 3281 | 5 | 12 | 2673 | 938 | 1589 |
| V1656 Cyg | 6.184 | 3281 | 6 | 8 | 3916 | 1124 | 1760 |
| V1657 Cyg | - | - | - | - | 2155 | 566 | 1002 |
| V1658 Cyg | 7.133 | 3281 | 5 | 7 | 2497 | 689 | 1233 |
| V1659 Cyg | 6.843 | 3281 | 5 | 12 | 3336 | 1014 | 1540 |
| V1660 Cyg | 6.236 | 3284 | 9 | 51 | 1734 | 506 | 970 |
| V1677 Cyg | 5.572 | 3282 | 14 | 43 | 5621 | 1761 | 2726 |
| V1678 Cyg | 0.919 | 5377 | 501 | 947 | 1914 | 318 | 468 |
| V1680 Cyg | 5.646 | 3284 | 9 | 42 | 2932 | 1014 | 1988 |

| Name | Bp-Rp | Teff (K) | Δ Teff (-) | Δ Teff (+) | Distance d (pc) | Δd (pc) (-) | Δd (pc) (+) |
|---|---|---|---|---|---|---|---|
| V1681 Cyg | 5.062 | 3325 | 46 | 87 | 4883 | 1537 | 2377 |
| V1682 Cyg | 6.672 | 3283 | 9 | 52 | 2574 | 719 | 1330 |
| V1683 Cyg | 1.834 | 4031 | 145 | 159 | 969 | 26 | 28 |
| V1684 Cyg | 4.353 | 3419 | 126 | 211 | 6237 | 1584 | 2303 |
| V1688 Cyg | - | - | - | - | 3112 | 987 | 1711 |
| V1689 Cyg | - | - | - | - | 5461 | 1617 | 2324 |
| V1690 Cyg | 6.405 | 3284 | 8 | 51 | 1184 | 197 | 292 |
| V1691 Cyg | - | - | - | - | 3071 | 927 | 1548 |
| V1692 Cyg | 4.993 | - | - | - | 1882 | 597 | 1224 |
| V1693 Cyg | 6.901 | 3281 | 5 | 8 | 2571 | 855 | 1582 |
| V1694 Cyg | 6.627 | 3283 | 9 | 52 | 2467 | 753 | 1425 |
| V2311 Cyg | 1.298 | 4897 | 188 | 196 | 1249 | 32 | 34 |

We searched also for photometric measurements and light curves analysis available from AAVSO and ASAS-SN databases. We analysed the valid photometric data using version 2.60 of the light curve and period analysis software PERANSO (Paunzen and Vanmunster, 2016). The analysis of the period was performed using several methods: Lomb-Scargle (Lomb 1976, Scargle 1982), ANOVA (Schwarzenberg-Czerny A., 1996) and FALC (Harris et al. 1989).

We chose these methods because they are powerful in detecting weak periodic signals, improving sensitivity of peak detection and damping alias periods.

The Lomb-Scargle and FALC are Fourier methods that attempt to represent a set of observations with a series of trigonometric functions (sines and cosines, with different periods, amplitudes, and phases). The Lomb-Scargle method transforms (unequally spaced) time series into a power spectrum, using a technique known as the Lomb periodogram.

The method was derived by Lomb in 1976, with improvements by Scargle in 1982. Although the Lomb-Scargle periodogram decomposes the data into a series of sine and cosine functions, it is similar to least-squares statistical methods, which minimises the difference between observed and modelled data. PERANSO uses the algorithm defined by Scargle but optimised using the Horne and Baliunas method (Horne & Baliunas 1986), which scales the periodogram by the total variance of the data, yielding a better estimation of the frequency of the periodic signal.

FALC is generally used for asteroid light curve period analysis but is a powerful tool for variable stars light curve studies as well. Currently, it is the only method in PERANSO that considers magnitude error values in the period determination.

ANOVA is a statistical method that instead of fitting the observation data with trigonometric functions, compares points in the observation data to other points at fixed time intervals or "lags" to see how different they are from one another. This method employs periodic orthogonal polynomials to fit observations, and the analysis of variance statistic to evaluate the quality of the fit and is suitable for the analysis of observation data that include non-sinusoidal periodic components.

For each value of the period, reported in this paper, a Fisher Randomization Test, with 200 permutations, was run with PERANSO software, in order to confirm the significance of the period we found. All results reported in this paper have a False Alarm Probabilities (FAP) less than 1%, indicating a high significance of the result. Any period with a FAP greater than 1% was disregarded. More precisely, PERANSO executes a permutation test or Monte Carlo Permutation Procedure (MCPP). Permutation tests are special cases of randomisation tests, i.e., tests that use randomly generated numbers for statistical inference. This test executes the selected period analysis calculation repeatedly, each time shuffling the magnitude values of the observations to form a new, randomised observation set, but keeping the observation times fixed (W. Press et al. 2007).

During our search for photometric data, we found for 23 variables a mismatch on the cross-reference of the Gaia DR2 source ID between the SIMBAD and ASAS-SN photometry database. The stars affected by this mismatch are summarised in Table 4. For some stars, the SIMBAD database does not report any Gaia DR2 source ID; for these cases, we have included in Table 4 the cross-reference source ID available from the Gaia Archive Search Tool based on the equatorial coordinates defined by SIMBAD. We investigated the reason for this difference: the result of our assessment is summarized in Table 4 and more details of our analysis are provided for the single stars in section 2.2.

Moreover, for each star, bibliographic references available from SIMBAD were reviewed.

Table 4         Potential misidentified variables

| Name | SIMBAD Gaia DR2 Source ID | ASAS-SN Gaia DR2 Source ID | Correct Source ID |
|---|---|---|---|
| NSV 13006 | 2061308294621233536 | 2061308294605585408 | SIMBAD - Note 1 |
| NSV 25055 | 2074619978810093952 | 2074619978810094080 | SIMBAD |
| NSV 25072 | 2062620664819487488 | 2062620870977919488 | Both incorrect - Note 2 |
| V433 Cyg | 2060941710574890368 | 2060941710574890496 | SIMBAD |
| V1622 Cyg | 2074170102454813056 | 2074170102454813184 | SIMBAD |
| V1623 Cyg | 2074087952635187840 | 2074087952635187712 | SIMBAD |
| V1625 Cyg | 2074497898646252672 | 2074497898646252544 | SIMBAD - Note 1 |
| V1628 Cyg | 2074538958561467008 | 2074538958561466880 | SIMBAD |
| V1630 Cyg | 2074515662659100800 | 2074515662659100672 | SIMBAD |
| V1632 Cyg | 2074517415005650944 | 2074517414985321984 | SIMBAD - Note 3 |
| V1633 Cyg | 2062369288973870976 | 20623692889738711 | SIMBAD |
| V1634 Cyg | 2074717423030592512 | 2074717423024363520 | SIMBAD |
| V1635 Cyg | 2061782665176034816 | 2061782665176035584 | SIMBAD - Note 4 |
| V1636 Cyg | 2074640560281165184 | 2074640560281165312 | SIMBAD |
| V1637 Cyg | 2062602557236419456 | 2062602557236419584 | SIMBAD |
| V1639 Cyg | 2061718996577182848 | 2061718996577182720 | SIMBAD |
| V1641 Cyg | 2061712193348895872 | 2061712193348896000 | Both incorrect - Note 5 |
| V1643 Cyg | 2060950884625235584 | 2060950884625235712 | SIMBAD |
| V1649 Cyg | 2062342728892073088 | 2062342728892072960 | SIMBAD |

| Name | SIMBAD Gaia DR2 Source ID | ASAS-SN Gaia DR2 Source ID | Correct Source ID |
|---|---|---|---|
| V1650 Cyg | 2061470128971060864 | 2061470128971060736 | SIMBAD |
| V1653 Cyg | 2061416012387664512 | 2061416012387664384 | SIMBAD |
| V1678 Cyg | 2061706455272623744 | 2061706455272623616 | SIMBAD |
| V1684 Cyg | 2068820948319278976 | 2068820948319279104 | SIMBAD |

Note 1: Gaia DR2 source ID is not defined in SIMBAD but derived from the Gaia Archive Search Tool.
Note 2: Correct source ID is defined in IBVS 6242 (Gaia DR2 source ID 2062620870978142592).
Note 3: Source ID is not defined in SIMBAD but derived by cross-reference with 2MASS J20120158+4100225.
Note 4: ID 2061782665176034816 is the most likely solution but ASAS.SN source ID cannot be excluded.
Note 5: Gaia G magnitude of both sources ID is fainter than expected from ASAS-SN V mean magnitude.

## 2.1. Variable stars in the Gaia Colour-Absolute Magnitude Diagram

To get a first general indication of the type of the variable stars listed in Table 1, we first analysed their position in a Gaia DR2 Colour-Absolute Magnitude Diagram (CAMD). Figure 1 shows $M_G$ vs. Bp-Rp for a training sample of about 166,000 variable stars. Different classes of variables are shown in different colours, each one occupying specific areas of the diagram (Jayasinghe et al. 2018b). In the same figure, we reported the position of the stars of IBVS 1302, as diamond-shaped dots. All photometric data reported in the diagram are not corrected for the interstellar and circumstellar extinction.

Figure 1 highlights that most of the variable stars of IBVS 1302 have photometric characteristics compatible with those of a Long Period Variable (LPV) of the training sample, i.e. belonging to the Mira, Semiregular or Irregular variable types.

Due to the overlap of the various groups of variables in the CAMD, this comparison does not allow a fine classification of the class of variables, but it is still a valid indicator of whether a variable is a candidate to be a Long Period Variable.

Based on their position in the Gaia DR2 CAMD, we have classified 56 out of 62 variables into the following groups, which are summarized in Table 5:

a. Not-LPV stars. It is a group of seven stars with colour index Bp-Rp < 1.9. Their nature is compatible with several groups of variables, but their photometric characteristics are not compatible with an LPV.
b. SR or YSO stars. It is a group of six stars which lay on the border between the Semiregular/Irregular variables and the Young Stellar Objects (YSO) groups. Their position in the CAMD follows the equation: $M_G = +0.65 \cdot (Bp-Rp) + 1.26$.
c. Probable YSO stars. Two stars, V1647 Cyg and V1692 Cyg, with colour index Bp-Rp > +3.5 and absolute magnitude $M_G > +6.6$ located in the low right corner of the CAMD. These stars have absolute magnitude not compatible with LPV stars and are likely young stellar objects.
d. LPV. These 41 stars show photometric characteristics compatible with a Long Period Variable.

For 6 of the 62 stars either the colour index Bp-Rp or the absolute magnitude $M_G$ is not available and therefore they cannot be classified with respect to the CAMD.

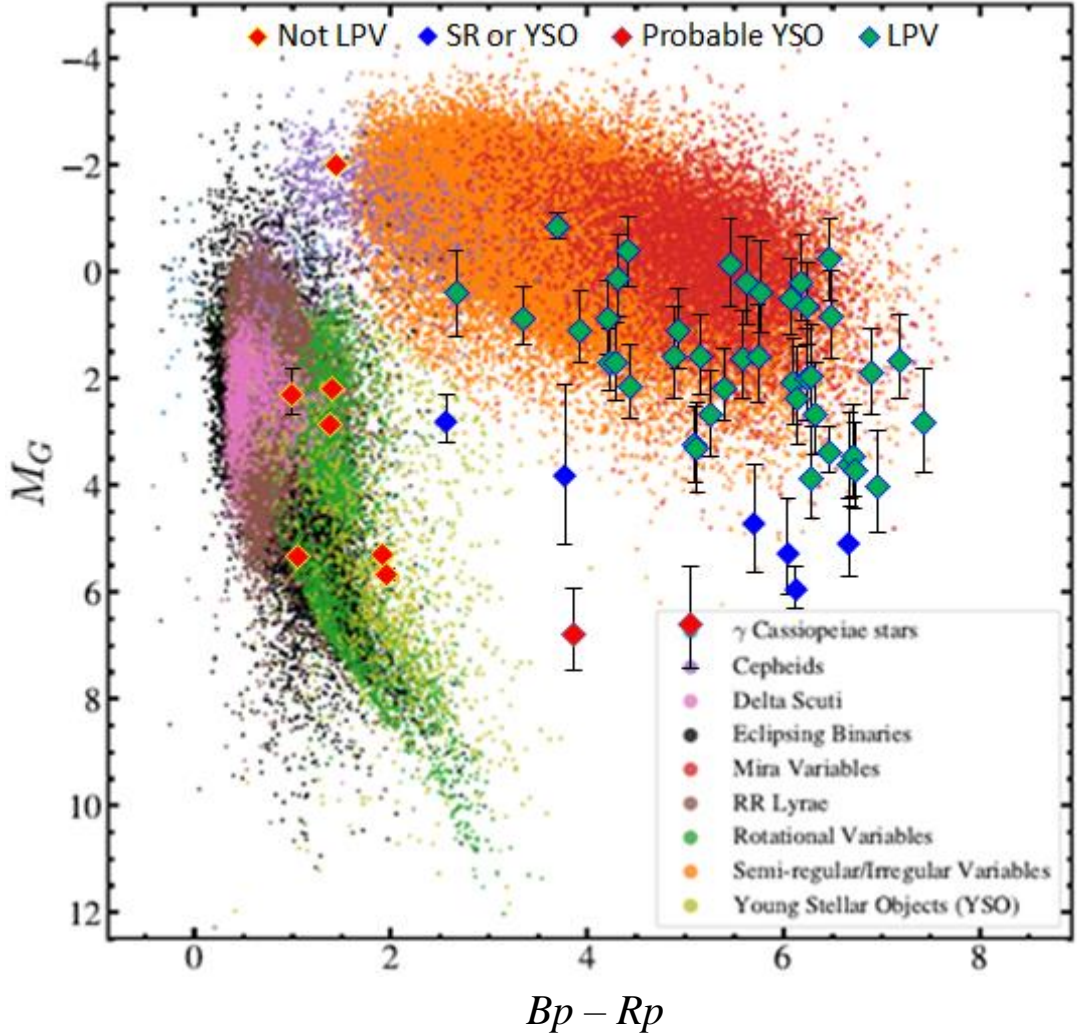

Figure 1　　IBVS 1302 variable stars in the Gaia DR2 CAMD

Table 5　　IBVS 1302 variable star classification based on Gaia DR2 CAMD

| Name | CAMD classification | Name | CAMD classification |
|---|---|---|---|
| V1641 Cyg | Not classified | V1629 Cyg | LPV |
| V1646 Cyg | Not classified | V1630 Cyg | LPV |
| V1657 Cyg | Not classified | V1631 Cyg | LPV |
| V1688 Cyg | Not classified | V1633 Cyg | LPV |
| V1689 Cyg | Not classified | V1634 Cyg | LPV |
| V1691 Cyg | Not classified | V1635 Cyg | LPV |
| NSV 25071 | Not LPV | V1636 Cyg | LPV |
| NSV 25072 | Not LPV | V1637 Cyg | LPV |
| NSV 25162 | Not LPV | V1638 Cyg | LPV |

| Name | CAMD classification | Name | CAMD classification |
|---|---|---|---|
| V1322 Cyg | Not LPV | V1639 Cyg | LPV |
| V1678 Cyg | Not LPV | V1640 Cyg | LPV |
| V1683 Cyg | Not LPV | V1642 Cyg | LPV |
| V2311 Cyg | Not LPV | V1643 Cyg | LPV |
| NSV 13006 | SR or YSO | V1645 Cyg | LPV |
| NSV 25117 | SR or YSO | V1648 Cyg | LPV |
| V1632 Cyg | SR or YSO | V1650 Cyg | LPV |
| V1649 Cyg | SR or YSO | V1651 Cyg | LPV |
| V1652 Cyg | SR or YSO | V1653 Cyg | LPV |
| V1680 Cyg | SR or YSO | V1654 Cyg | LPV |
| V1647 Cyg | Probable YSO | V1655 Cyg | LPV |
| V1692 Cyg | Probable YSO | V1656 Cyg | LPV |
| NSV 25019 | LPV | V1658 Cyg | LPV |
| NSV 25055 | LPV | V1659 Cyg | LPV |
| NSV 25113 | LPV | V1660 Cyg | LPV |
| V0433 Cyg | LPV | V1677 Cyg | LPV |
| V1622 Cyg | LPV | V1681 Cyg | LPV |
| V1623 Cyg | LPV | V1682 Cyg | LPV |
| V1625 Cyg | LPV | V1684 Cyg | LPV |
| V1626 Cyg | LPV | V1690 Cyg | LPV |
| V1627 Cyg | LPV | V1693 Cyg | LPV |
| V1628 Cyg | LPV | V1694 Cyg | LPV |

## 2.2. Detailed data analysis

Details of the relevant information available from the bibliographic references, catalogues and databases and the results of our period analysis and assessment are reported in this section for each star.

In our analysis, we adopted the following relations between the Gaia DR2 (filter G), ASAS-SN (filter V) and the original mean infrared photographic magnitude ($I_n$):

$$I_n - 0.1 \leq G \leq I_n + 2.2 \qquad [1]$$
$$G \leq V \leq G + 1.4 \qquad [2]$$

Equation [1] is derived from the G-$I_n$ range of 60 stars of IBVS 1302.

Equation [2] is derived from the G-V = f(V-I) colour-colour transformation defined for Gaia DR2 photometry (Evans et al. 2018) and considering $I_n$ - I = 0.45.

We used common sections and/or figures and tables when similar characteristics apply to more than one variable.

**NSV 13006**

The equatorial coordinates of this star were revised by IBVS 6242 (Nesci, 2018), as R.A. 20h 19m 03.95s and Decl. +38° 54' 45.7" and cross-matched with Gaia DR2 source ID 2061308294621233536.

Its colour index Bp-Rp = 2.493 and absolute magnitude $M_G$ = 2.76 (-0.49, +0.42) are compatible with a variable at the border between the Semiregular/Irregular and YSO groups of Figure 1.

At a distance of 10 arcsec from the revised equatorial coordinates, the ASAS-SN Catalog of Variable Stars II (Jayasinghe et al. 2018b) reports a generic variable, with an amplitude of 0.18 mag and a period of 339 days. The catalogue identifies this star, ASASSN-V J201904.15+385435.9, with a different Gaia DR2 source ID 2061308294605585408. The ASAS-SN cross-reference ID is deemed incorrect because the apparent magnitude G = 19.76 associated to this Gaia source ID is not consistent to the mean V magnitude = 10.63 measured by the survey and to the original 14.4 ÷ 15.6 magnitude range observed in the infrared photographic band. Based on the observed infrared magnitude, the most likely correct reference for the variable observed by ASAS-SN is the 11$^{th}$ magnitude star HD 228883 (Gaia DR2 2061308294621232384). For HD 228883, based on 658 measurements of ASAS-SN database covering 1377 days and applying the Lomb-Scargle method, we found a weak variability, with a mean fit curve amplitude of 0.03 mag, a period of 192 ± 12 days (Figure 2), and a maximum at epoch 2458322 ± 12 HJD.

No photometric data were available for Gaia DR2 source ID 2061308294621233536 from ASAS-SN database (Kochanek et al. 2017) because this is a 16$^{th}$ magnitude star in the G band, fainter than the V ~15 limit of the survey, therefore no further analysis was possible.

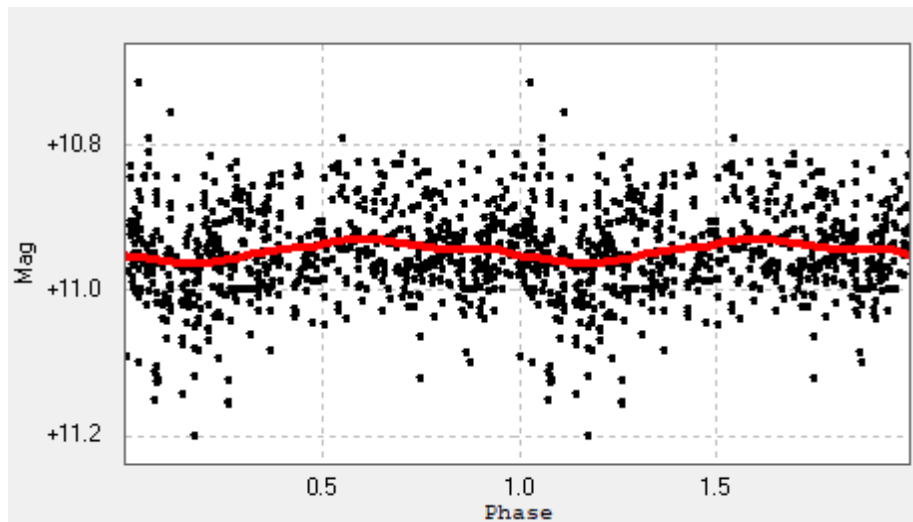

Figure 2    HD 228883 - V mag vs Phase (192 ± 12 days period - Lomb-Scargle)

**NSV 25019**

The colour index Bp-Rp = 4.259 and the absolute magnitude $M_G$ = 0.09 (-0.80, +0.72) are compatible with a LPV star (see Figure 1).

No light curve, type or period estimation is available from ASAS-SN photometry database or Gaia DR2 database.

Our period analysis, based on 655 ASAS-SN observations in the filter V on a time span of 1377 days, applying the Lomb-Scargle did not highlight a reliable periodicity for this variable.
The photometric data confirm the uncertain original classification of a slow irregular variable star.

**NSV 25055**

This suspected variable is identified as Gaia DR2 source ID 2074619978810093952.
The colour index Bp-Rp = 5.521 and the absolute magnitude $M_G$ = 1.61 (-0.92, +0.72) are compatible with a LPV star.
The ASAS-SN Catalog of Variable Stars II reports this variable as Semiregular, with a mean magnitude V = 17.62, an amplitude of 0.5 mag and a period of 7.6486 days. The catalogue identifies this star, ASASSN-V J201210.10+412224.0, with a different Gaia DR2 source ID 2074619978810094080. The ASAS-SN cross-reference ID is deemed incorrect because we could not find this source within a circle of 60" centred on NSV 25055 equatorial coordinates. We also observed that the mean magnitude in V, reported by ASAS-SN, is not compatible with the original 12.4 ÷ 13.5 infrared magnitude range, whilst is consistent with the $14^{th}$ magnitude in the filter G of Gaia DR2 source ID 2074619978810093952.
We did not perform any period analysis on this star because not sufficient valid data were available from the ASAS-SN photometry database.

**NSV 25071**

This star is identified as Gaia DR2 source ID 2062218896403414784. The colour index Bp-Rp = 1.326 and the absolute magnitude $M_G$ = -2.04 ± 0.04 are not compatible with a LPV star.
The ASAS-SN Catalog of Variable Stars II confirms the uncertain original eclipsing nature of this variable and classifies it as an Algol eclipse binary, with an amplitude of 0.67 mag and a period of 8.8075552 days.

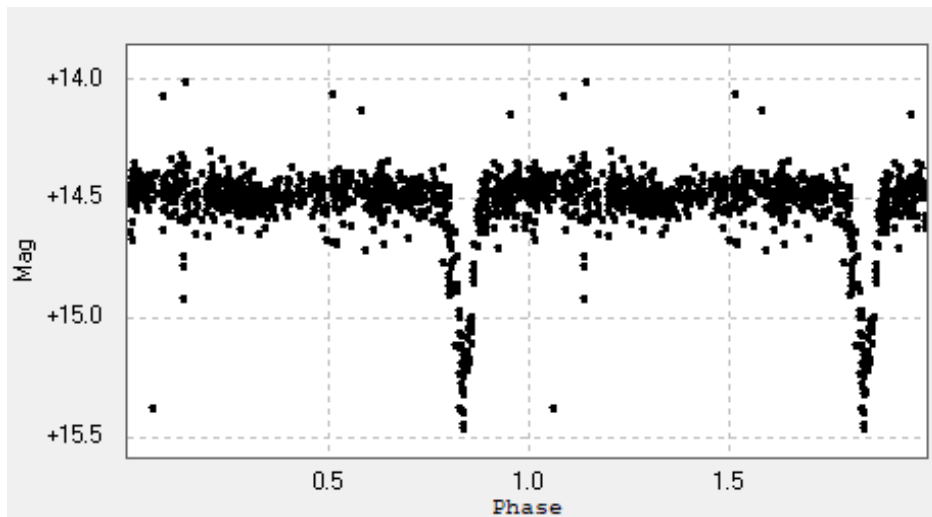

Figure 3    NSV 25071 - V mag vs Phase (8.8079 ± 0.0027 days period - ANOVA)

Our light curve analysis confirmed the Algol nature of this variable. We applied Lomb-Scargle, ANOVA and FALC methods on 871 ASAS-SN observations in the filter V on a time span of 1377 days. The weighted average of the periods we obtained with the three methods is: 8.8079 ± 0.0027 (Figure 3), that confirms the ASAS-SN result.

We identified a minimum of the light curve at 2457710.698 ± 0.019 HJD.

**NSV 25072**

The equatorial coordinates of this star were revised by IBVS 6242 (Nesci, 2018), as R.A. 20h 15m 07.07s and Decl. +41° 17' 47.5 and cross-matched to Gaia DR2 source ID 2062620870978142592. This source is a 14$^{th}$ magnitude star, whose apparent luminosity is consistent with the observed original infrared 12.8 ÷ 13.8 magnitude range.

Due to the initial wrong identification, the SIMBAD database reports an incorrect cross-reference to GAIA DR2 source ID 2062620664819487488.

The variable is a quite close system 498 (-8, +9) pc away with a temperature of 5332 (-80, +18) K. Its colour index Bp-Rp = 0.978 and absolute magnitude $M_G$ = 5.32 ± 0.04 exclude that this star could be a LPV star. The original uncertain Cepheid type is not compatible with the calculated absolute magnitude and is therefore very unlikely.

The ASAS-SN Catalog of Variable Stars II identifies this star, ASASSN-V J201507.25+411751.7, as a rotational variable, with a mean magnitude in V = 8.35, with no period defined and an amplitude of 0.26 mag. It reports an incorrect cross-reference to Gaia DR2 source ID 2062620870977919488, which is a 16$^{th}$ magnitude star. The 8$^{th}$ magnitude rotational variable, ASASSN-V J201507.25+411751.7 source, identified by ASAS-SN, is the double star HD 192803. We performed a period analysis of HD 192803 using 675 valid observations available from ASAS-SN in the V filter and covering a time span of 1358 days. Applying the Lomb-Scargle algorithm, we found a period of 29.6 ± 0.2 hours (Figure 4).

We did not find any other photometric data for fainter stars close to HD 192803, so no further analysis of NSV 25072 was possible.

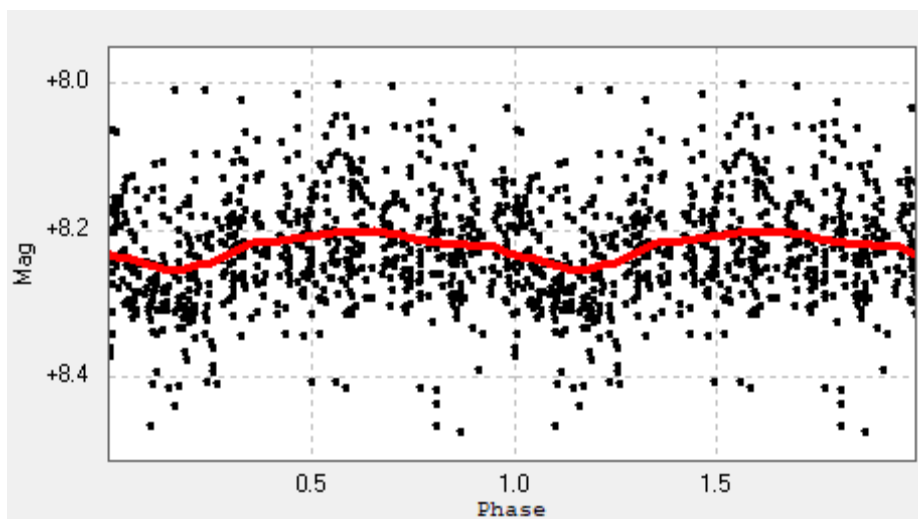

Figure 4        HD 192803 - V mag vs Phase (29.6 ± 0.2 hours period - Lomb-Scargle)

**NSV 25113**

The equatorial coordinates of this star were revised by IBVS 6242 (Nesci, 2018), as R.A. 20h 21m 18.81s and Decl. +39° 03' 05.4" and cross-matched to Gaia DR2 source ID 2061392957003016704. The colour index Bp-Rp = 5.042 and the absolute magnitude $M_G$ = 3.22 (-0.73, +0.72) are compatible with a LPV star. There is not a light curve computed for this variable by ASAS-SN photometry database; using as input the revised equatorial coordinates the database returns the photometry of a V = 13.6 star, that we could not identify.

The period analysis we performed on this star highlighted a weak variability, with a period of 250 ± 17 days and a mean fit curve amplitude of 0.04 mag in the V filter (Figure 5).

The mean magnitude and amplitude that can be derived by ASAS-SN observations do not match with the original infrared 14.4 ÷ 15.3 magnitude range and uncertain type, so we could exclude that these data apply to NSV 25113. We note that the AAVSO database classifies this variable as an uncertain Mira type, but we did not identify any reference paper for this classification.

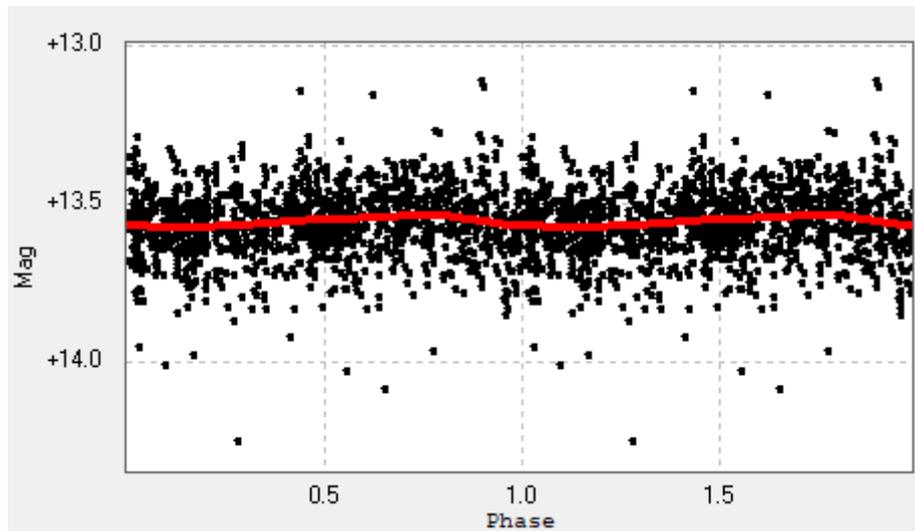

Figure 5    Unidentified star - V mag vs Phase (250 ± 17 days period - Lomb-Scargle)

**NSV 25117**

The type for this variable was not defined in the original work. This variable is cross-referenced to Gaia DR2 source ID 2068494908757889024. Based on Gaia DR2 data, the distance of this star is 2080 (-630, +1251) pc, with an effective temperature of 3284 (-8, +39) K, typical of a late spectral type. Its colour index Bp-Rp = 5.985 and absolute magnitude $M_G$ = 5.25 (-1.02, +0.78) locate this star on the border between the Semiregular/Irregular variables and the YSO groups of Figure 1. No light curve could be determined because valid measurements were not available from ASAS-SN database.

**NSV 25162**

No variable type and/or period is reported in the original work for this suspected variable, that is associated to Gaia DR2 source ID 2064177229676836608 by SIMBAD and Gaia DR2 databases. Around the equatorial coordinates of this source, the ASAS-SN database contains only the photometry of Gaia DR2 source ID 2064177264036574720, a star with mean magnitude V = 12.43, 9 arcsec from NSV 25162. There is no photometry of the suspected variable available in the ASAS-SN database. Based on its colour index Bp-Rp = 1.887 and its absolute magnitude $M_G$ = 5.65 (-0.13, +0.12) we could establish that this suspected variable is not a LPV star.

**V433 Cyg**

This variable is an M.6.5 spectral type star, identified as Gaia DR2 source ID 2060941710574890368 by SIMBAD database. Its colour index Bp-Rp = 5.710 and absolute magnitude $M_G$ = 0.36 (-0.98, +0.75) are compatible with a LPV star. The Gaia DR2 database classifies this star as LPV candidate, with a period of 384 ± 33 days (Mowlavi, 2018).

The ASAS-SN Catalog of Variable Stars II identifies this star, ASASSN-V J201539.86+382545.2, with a different Gaia DR2 source ID 2060941710574890496. The ASAS-SN cross-reference ID is deemed incorrect because this source cannot be found in a radius of 60 arcsec, and photometric data reported in the database are those of source ID 2060941710574890368. The ASAS-SN database classifies V433 Cyg as Mira variable, with a mean magnitude amplitude V = 14.8, an amplitude of 2.17 mag and a period of 369 days.

An unreferenced Mira type with a period of 417 days is also reported by AAVSO database.

Our period analysis, based on 625 ASAS-SN observations in the filter V on a time span of 1374 days, applying the Lomb-Scargle method, highlighted a period of 368 ± 33 days (Figure 6), with a maximum at epoch 2457490 ± 6 HJD. We also noted that the original 8.7 ÷ 11.4 infrared magnitude range, computed using equations [1] and [2], shows a higher level of luminosity compared to the expected mean V = 13.6 mag that has been measured by ASAS-SN. Nevertheless, we did not identify any alternative source with such magnitude in a radius of 60 arcsec and we believe that a misidentification of a star is very unlikely. The difference may have different explanations: invalidity of equations [1] and [2], overestimation of original magnitude, or variation of the star mean level of luminosity. The original Mira type classification is confirmed whilst the period is likely shorter than the original and AAVSO values.

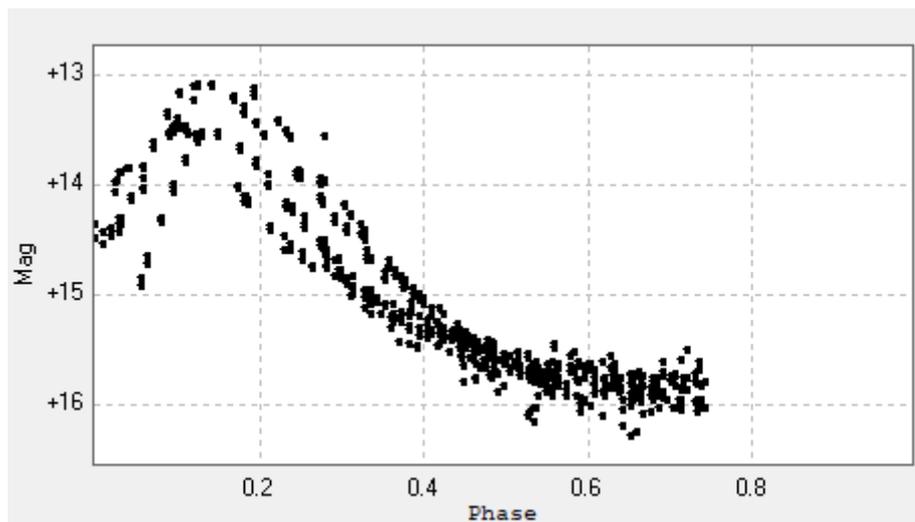

Figure 6    V433 Cyg - V mag vs Phase (368 ± 33 days period - Lomb-Scargle)

**V1322 Cyg**

This star is 1461 (-51, +55) pc away and has a temperature of 4828 (-143, +136) K. This value of the effective temperature available from Gaia DR2 is not consistent with its spectral type classification B0.2IIIe (Laur, et al. 2017). Its colour index Bp-Rp = 1.371 and absolute magnitude $M_G$ = -2.04 ± 0.08 are not compatible with a LPV star.

The ASAS-SN Catalog of Variable Stars II classifies this variable as an eruptive γ Cas (GCAS), with a mean magnitude V = 9.51 and an amplitude of 0.39 mag. This classification is confirmed by the observations of the Optical Monitoring Camera (OMC) onboard INTEGRAL (Alfonso-Garzón et al. 2012), which provides a V = 9.15 ÷ 9.31 magnitude range.

**V1622 Cyg**

This star is identified as Gaia DR2 source ID 2074170102454813056 by SIMBAD database.

The colour index Bp-Rp = 5.394 and the absolute magnitude $M_G$ = -0.17 (-0.84, +0.78) are compatible with a LPV star.

The Gaia DR2 database classifies this star as LPV candidate, with a period of 354 ± 32 days.

The ASAS-SN Catalog of Variable Stars II identifies this star, ASASSN-V J200732.14+404345.6, with a different Gaia DR2 source ID 2074170102454813184 and classifies this variable as Semiregular, with a mean magnitude V = 16.01, an amplitude of 1.41 mag and a period of 369 days. The ASAS-SN cross-reference ID is deemed incorrect because this source cannot be found in a radius of 60 arcsec and photometric data reported in the database are those of source ID 2074170102454813056.

We applied the ANOVA method on 152 ASAS-SN observations in the filter V on a time span of 1346 days and highlighted a period of 366 ± 13 days (Figure 7), with a mean fit curve amplitude of 0.98 mag and a maximum at epoch 2457825 ± 10 HJD.

The original Mira type classification is not confirmed by ASAS-SN analysis, whilst consistent values for the period around one year are found.

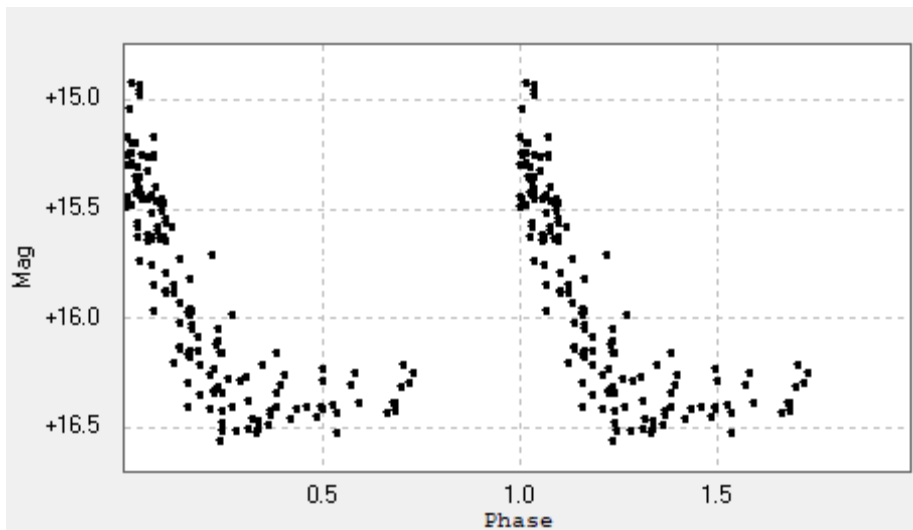

Figure 7    V1622 Cyg - V mag vs Phase (366 ± 13 days period - ANOVA)

**V1623 Cyg**

This star is identified as Gaia DR2 source ID 2074087952635187840 by SIMBAD database. The colour index Bp-Rp = 5.332 and the absolute magnitude $M_G$ = 2.17 (-0.77, +0.59) are compatible with a LPV star.

The Gaia DR2 database classifies this star as LPV candidate, with a period of 306 ± 22 days.

The ASAS-SN Catalog of Variable Stars II identifies this star, ASASSN-V J200737.05+395635.3, with a different Gaia DR2 source ID 2074087952635187712 and classifies this variable as Semiregular, with a mean magnitude V = 16.22, an amplitude of 1.93 mag and a period of 310 days. The ASAS-SN cross-reference ID is deemed incorrect because this source cannot be found in a radius of 60 arcsec and photometric data reported in the database are those of source ID 2074087952635187840.

Our period analysis, based on 210 ASAS-SN observations in the filter V on a time span of 1332 days, applying the Lomb-Scargle method highlighted three potential shorter periods at 103 ± 3 days (Figure 8), 153 ± 10 days (Figure 9) and 222 ± 29 days (Figure 10). The mean fit curve amplitude is in the range 0.78 ÷ 0.92 mag. A maximum was found at epoch 2458384 ± 2 HJD. From a statistical point of view all the three periods are possible but the period at 103 days is the most probable due to its better accuracy.

We searched for periods around 300 days as reported by Gaia DR2 and ASAS-SN; using all three methods described in section 2 but we did not find any reliable solution. As an example, Figure 11 shows the results of the period search performed with the Lomb-Scargle method, where the low value of theta parameter shows that no valid solutions are present around 300 days.

The original Mira type classification is not confirmed by ASAS-SN analysis and a shorter period of 103 days has been identified.

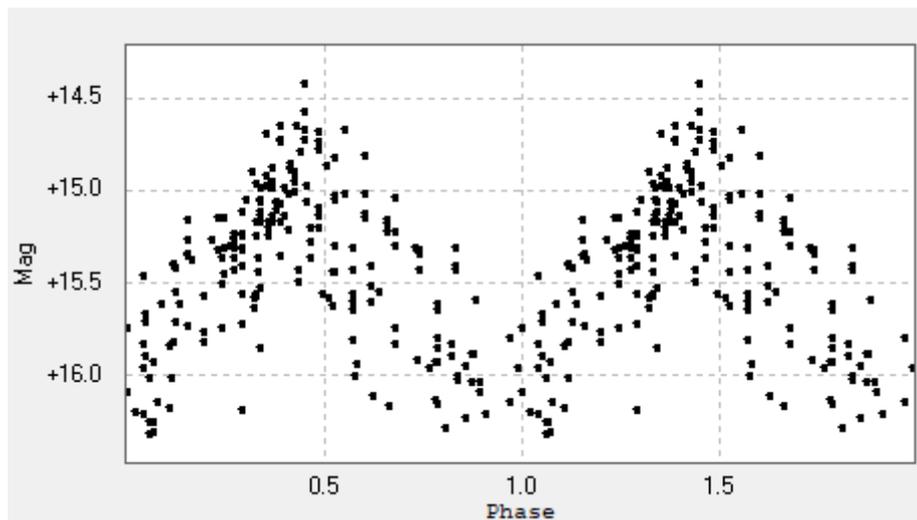

Figure 8    V1623 Cyg - V mag vs Phase (103 ± 3 days period - Lomb-Scargle)

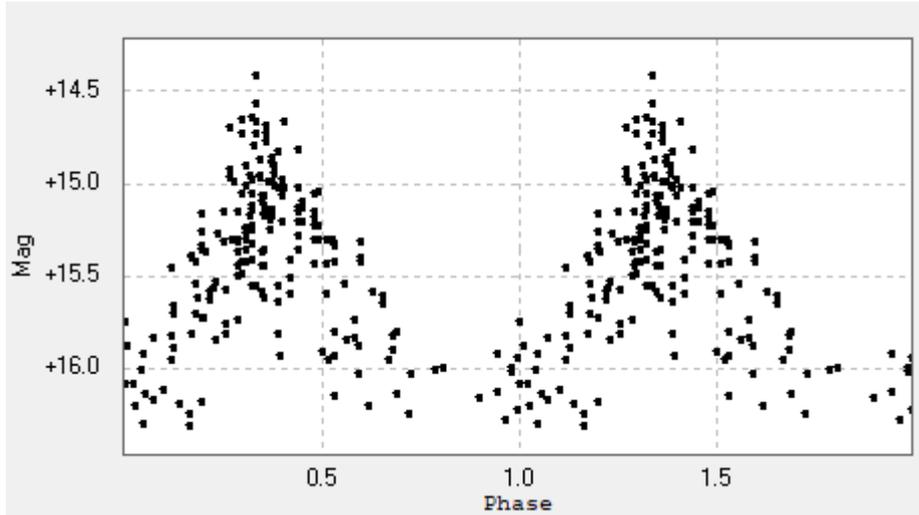

Figure 9  V1623 Cyg - V mag vs Phase (153 ± 10 days period - Lomb-Scargle)

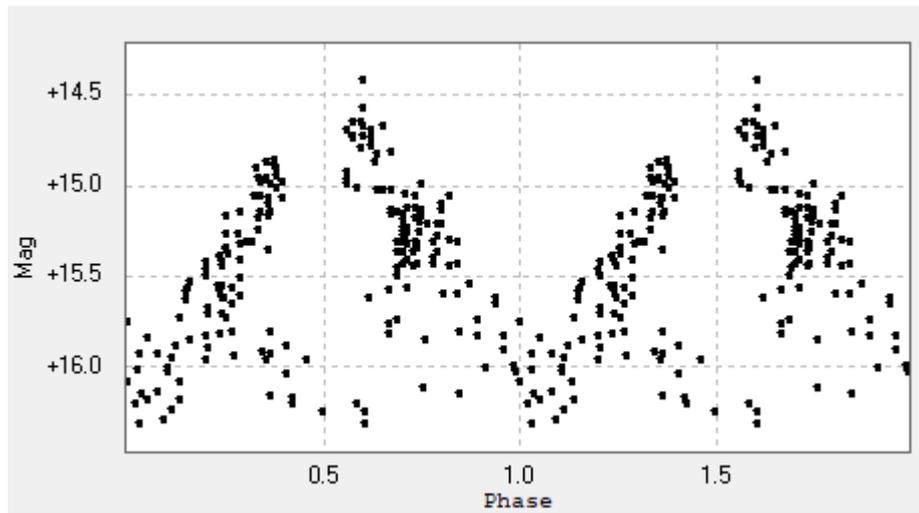

Figure 10  V1623 Cyg - V mag vs Phase (222 ± 29 days period - Lomb-Scargle)

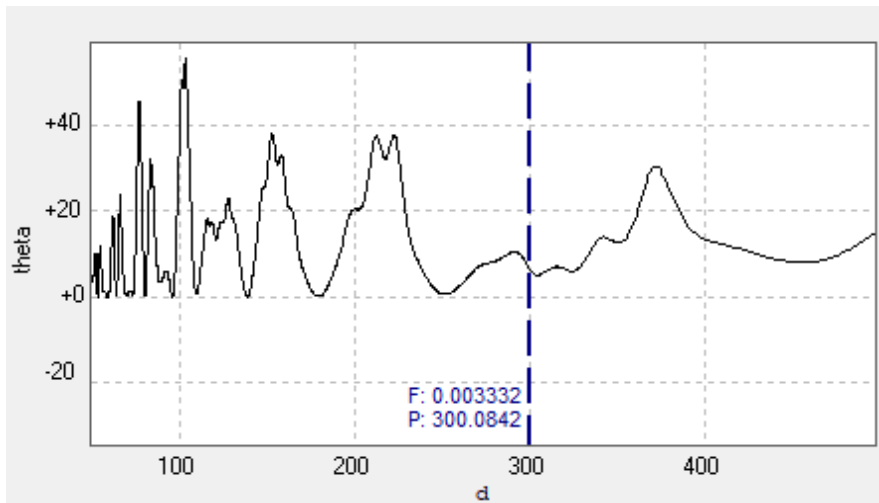

Figure 11  V1623 Cyg - Period analysis in the range 50 ÷ 500 days (Lomb-Scargle)

**V1625 Cyg**

This star is identified as Gaia DR2 source ID 2074497898646252672 by SIMBAD database. The colour index Bp-Rp = 3.865 and the absolute magnitude $M_G$ = 1.09 (-0.74, +0.60) are compatible with a LPV star.

The ASAS-SN Catalog of Variable Stars II identifies this star, ASASSN-V J200926.04+404347.2, with a different Gaia DR2 source ID 2074497898646252544 and classifies this variable as Semiregular, with a mean magnitude V = 15.61, an amplitude of 0.77 mag and a period of 480 days.

The ASAS-SN cross-reference ID is deemed incorrect because this source cannot be found in a radius of 60 arcsec and photometric data reported in the database are those of source ID 2074497898646252672.

Our period analysis, based on 238 ASAS-SN observations in the filter V on a time span of 1321 days, applying the ANOVA method highlighted a period of 464 ± 30 days (Figure 12), with a mean fit curve amplitude of 0.63 mag and a maximum at epoch 2457669 ± 5 HJD.

The original uncertain Semiregular type classification is confirmed by ASAS-SN analysis.

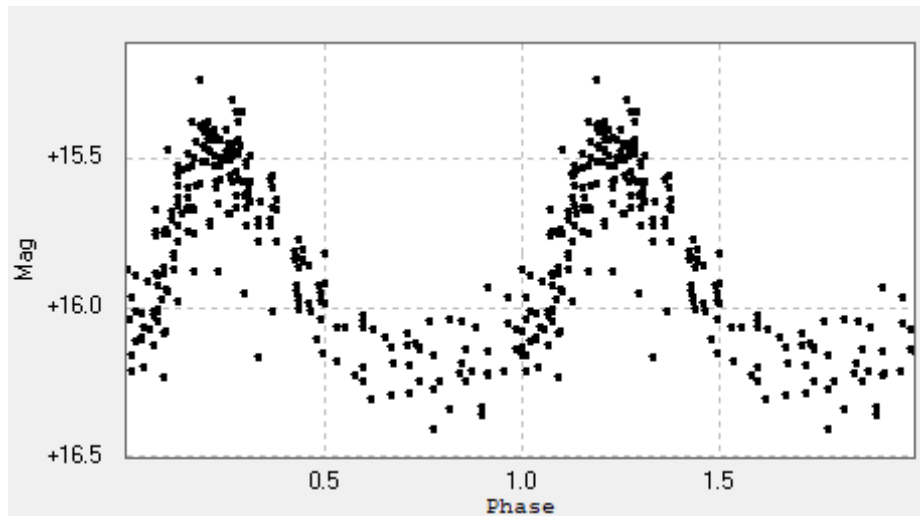

Figure 12    V1625 Cyg - V mag vs Phase (464 ± 30 days period - ANOVA)

**V1626 Cyg**

The colour index Bp-Rp = 4.822 and the absolute magnitude $M_G$ = 1.55 (-0.94, +0.81) are compatible with a LPV star.

The Gaia DR2 database classifies this star as LPV candidate, with a period of 626 ± 151 days.

The ASAS-SN Catalog of Variable Stars II classifies this variable as Semiregular, with a mean magnitude V = 16.05, an amplitude of 0.35 mag and a period of 482 days.

Our period analysis, based on 95 ASAS-SN observations in the filter V on a time span of 1322 days, applying the Lomb-Scargle method did not highlight a reliable periodicity for this variable.

The original uncertain Semiregular type classification is confirmed by ASAS-SN analysis with a shorter period.

**V1627 Cyg**

The Gaia DR2 database classifies this star as LPV candidate, with a period of 310 ± 36 days. The membership of this star to the LPV group is also confirmed by its colour index Bp-Rp = 3.285 and absolute magnitude $M_G$ = 0.85 (-0.59, +0.50).

The ASAS-SN Catalog of Variable Stars II reports this variable as Semiregular, with a mean magnitude V = 14.36, an amplitude of 0.34 mag and a period of 299 days.

Our period analysis, with Lomb-Scargle algorithm, of 667 valid observations available from ASAS-SN in the V filter and covering a time span of 1358 days, highlighted a period of 295 ± 11 days (Figure 13) and a maximum at epoch 2457929 ± 3 HJD.

The membership of V1627 Cyg to the LPV group and the original period are confirmed.

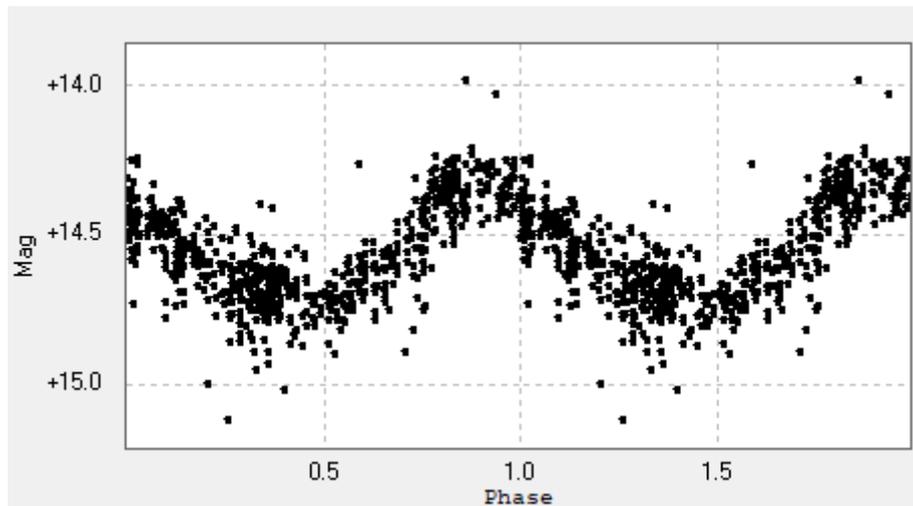

Figure 13    V1627 Cyg - V mag vs Phase (295 ± 11 days period - Lomb-Scargle)

**V1628 Cyg**

This star is identified as Gaia DR2 source ID 2074538958561467008 by SIMBAD database.

The Gaia DR2 database classifies this star as LPV candidate, with a period of 258 ± 13 days. The membership of this star to the LPV group is also confirmed by the colour index Bp-Rp = 4.870 and the absolute magnitude $M_G$ = 1.06 (-0.76, +0.74).

The ASAS-SN Catalog of Variable Stars II identifies this star, ASASSN-V J201051.40+411555.0, with a different Gaia DR2 source ID 2074538958561466880 and classifies this variable as Semiregular, with a mean magnitude V = 16.27, an amplitude of 1.25 mag and a period of 257 days. The ASAS-SN cross-reference ID is deemed incorrect because this source cannot be found in a radius of 60 arcsec and photometric data reported in the database are those of source ID 2074538958561467008.

Our period analysis, based on 225 valid observations available from ASAS-SN in the V filter and covering a time span of 1300 days, applying the Lomb-Scargle algorithm, highlighted a period of 219 ± 37 days (Figure 14) and a maximum at epoch 2457943 ± 2 HJD. The original classification and period of V1628 Cyg are confirmed.

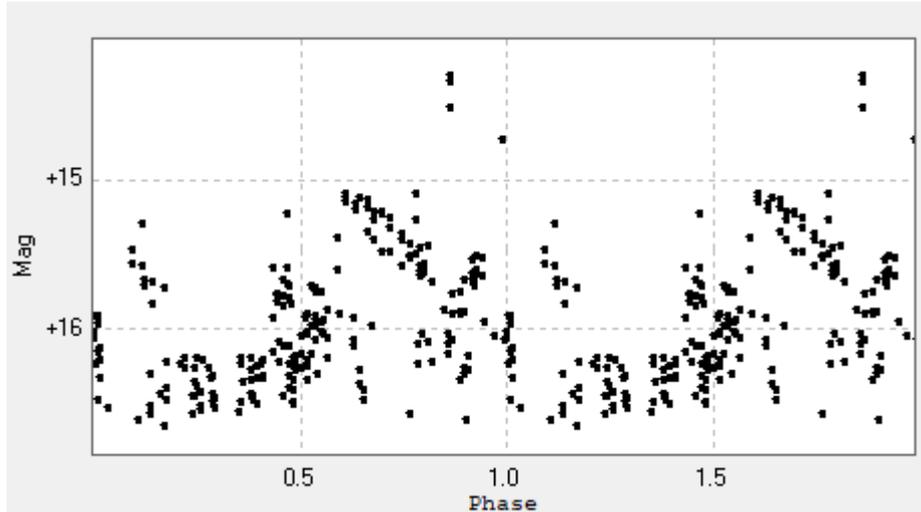

Figure 14    V1628 Cyg - V mag vs Phase (219 ± 37 days period - Lomb-Scargle)

**V1629 Cyg**

The colour index Bp-Rp = 4.230 and the absolute magnitude $M_G$ = 1.68 (-0.76, +0.70) are compatible with a LPV star. The ASAS-SN Catalog of Variable Stars II classifies this variable as Semiregular, with a mean magnitude V = 16.0, an amplitude of 0.35 mag and a period of 503 days. Our period analysis, based on 274 ASAS-SN observations in the filter V on a time span of 1327 days, applying the Lomb-Scargle method, highlighted a period of 476 ± 67 days (Figure 15), with a mean fit curve amplitude of 0.21 mag and a maximum at epoch 2457186 ± 6 HJD. The original Semiregular type and the uncertain value of the period are confirmed.

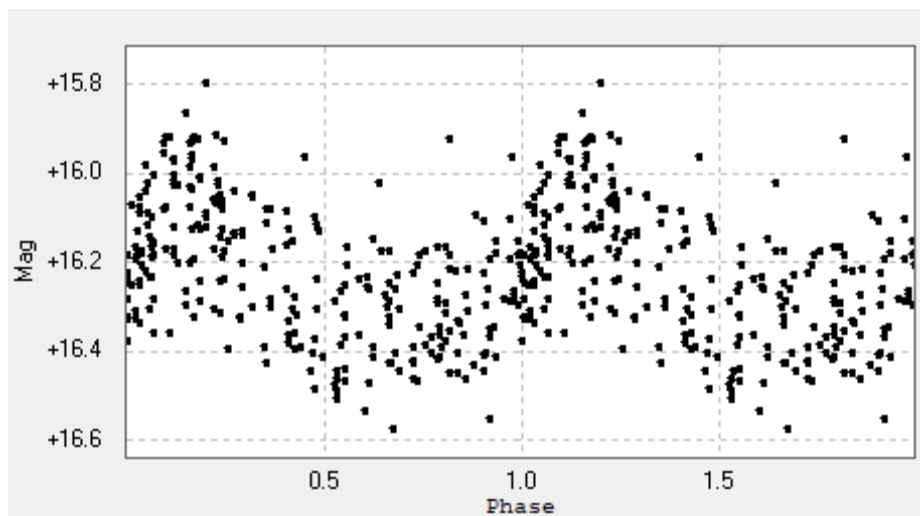

Figure 15    V1629 Cyg - V mag vs Phase (476 ± 67 days period - Lomb-Scargle)

**V1630 Cyg**

This star is identified as Gaia DR2 source ID 2074515662659100800 by SIMBAD database. The Gaia DR2 database classifies this star as LPV candidate, with a period of 416 ± 76 days. The membership of this star to the LPV group is also confirmed by the colour index Bp-Rp = 4.157 and the absolute magnitude $M_G$ = 1.68 (-0.62, +0.51).

The ASAS-SN Catalog of Variable Stars II identifies this star, ASASSN-V J201116.24+410435.0, with a different Gaia DR2 source ID 2074515662659100672 and classifies this variable as Semiregular, with a mean magnitude V = 16.83, an amplitude of 1.17 mag and a period of 215 days. The ASAS-SN cross-reference ID is deemed incorrect because this source cannot be found in a radius of 60 arcsec. It is also noted that the mean V magnitude reported by ASAS-SN is fainter than expected from equations [1] and [2], which foresee a magnitude V in the range 14.1 ÷ 15.6. We did not perform any analysis of the light curve because not sufficient valid measurements were available from ASAS-SN photometry database. The Mira classification type and period of 415 days are compatible with the Gaia DR2 analysis but are not consistent with the ASAS-SN classification.

**V1631 Cyg**

The colour index Bp-Rp = 4.376 and the absolute magnitude $M_G$ = 2.12 (-0.79, +0.62) are compatible with a LPV star.

The ASAS-SN Catalog of Variable Stars II reports this variable as Semiregular, with a mean magnitude V = 16.75, an amplitude of 1.46 mag and a period of 497 days.

We performed a period analysis using 141 valid observations available from ASAS-SN in the V filter and covering a time span of 1330 days. Applying the Lomb-Scargle algorithm, we found a period of 235 ± 16 days (Figure 16) and a maximum at epoch 2458030 ± 6 HJD. Our analysis of a longer period, in accordance with ASAS-SN result, did not identify any reliable solution. Figure 17 shows a potential period, statistically valid but with an extremely poor accuracy at 543 ± 105 days.

The Mira type is not confirmed by the ASAS-SN result and period could be significantly shorter compared to the original classification.

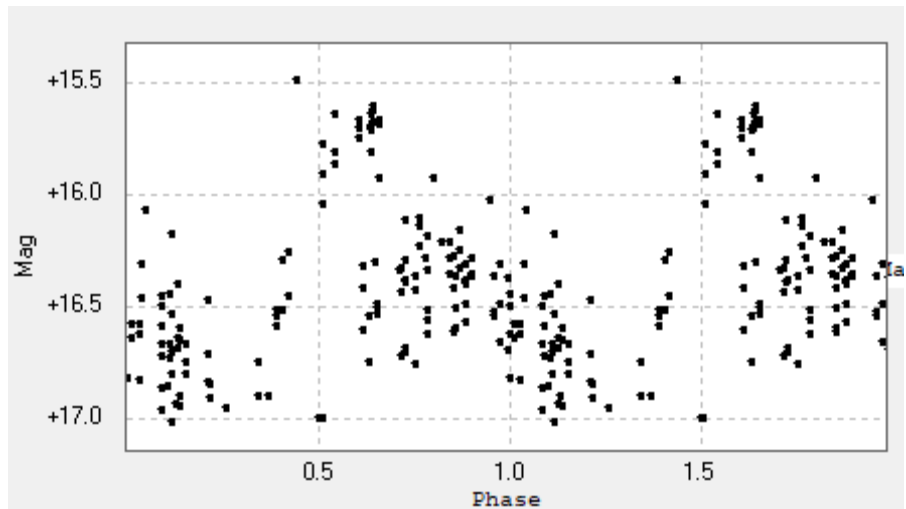

Figure 16      V1631 Cyg - V mag vs Phase (235 ± 16 days period - Lomb-Scargle)

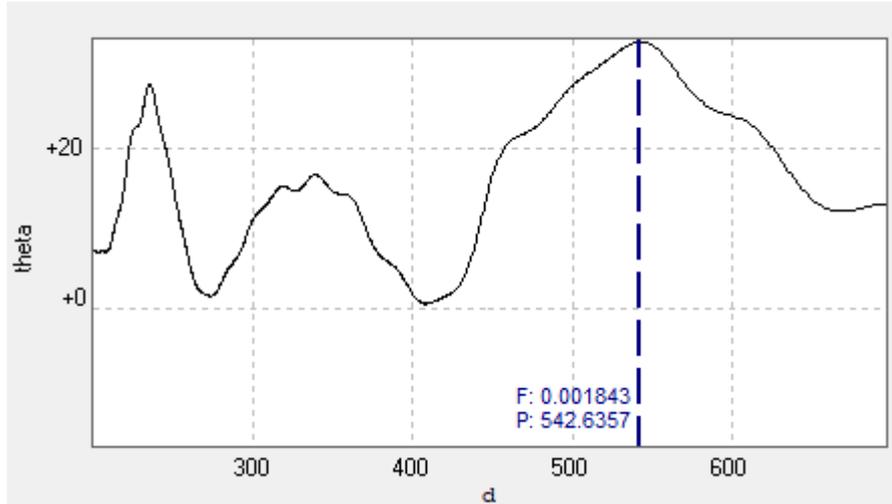

Figure 17    V1631 Cyg - Period analysis in the range 200 ÷ 700 days (Lomb-Scargle)

**V1632 Cyg**

This star is identified as 2MASS J20120158+4100225 by the SIMBAD database. Based on the equatorial coordinates available from the 2MASS catalogue we associated this star to Gaia DR2 source ID 2074517415005650944, with magnitude G = 15.307.

Based on its colour index Bp-Rp = 3.703 and absolute magnitude $M_G$ = 3.79 (-1.70, +1.29), this variable is compatible with a star on the border between the Semiregular/Irregular variables and the YSO groups of CAMD in Figure 1.

The ASAS-SN Catalog of Variable Stars II reports this star, ASASSN-V J201201.61+410021.3, as a generic variable, with a mean magnitude V = 15.42, an amplitude of 0.46 mag and a period of 444 days. The ASAS-SN photometry database identifies this variable with a different Gaia DR2 source ID 2074517414985321984, that is a star with magnitude G = 16.578, separated by 0.9 arcsec from source ID 2074517415005650944.

Even if, based on Gaia DR2 data, this star is a remarkably close double star, in which at least one of the components is variable, we believe the ASAS-SN cross-reference to Gaia DR2 source ID 2074517414985321984 is incorrect because the magnitude of this star does not match with the mean magnitude observed by ASAS-SN and falls outside the magnitude range G = 13.4 ÷ 15.7 expected from the original infrared measurements and derived from equation [1]

The period analysis we performed using 633 valid observations available from ASAS-SN in the V filter and covering a time span of 1358 days and applying the Lomb-Scargle algorithm, highlighted a period of 160 ± 7 days (Figure 18), with a maximum at epoch 2458250 ± 8 HJD. Our analysis of a longer period, in accordance with the original and ASAS-SN results, found also a second potential period at 425 ± 58 days but with a greater uncertainty (Figure 19).

The original Semiregular type and the period of 410 days defined for V1632 Cyg are not confirmed and may require a revision.

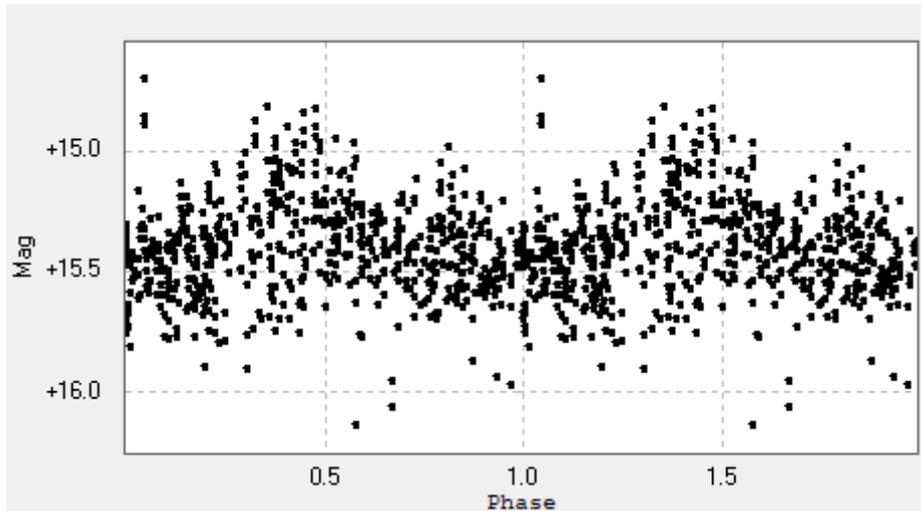

Figure 18    V1632 Cyg - V mag vs Phase (160 ± 7 days period - Lomb-Scargle)

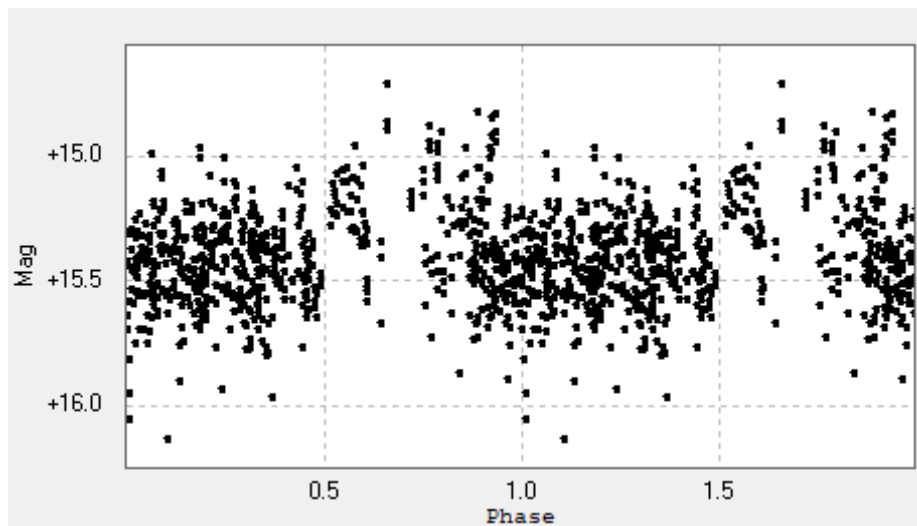

Figure 19    V1632 Cyg - V mag vs Phase (425 ± 58 days period - Lomb-Scargle)

**V1633 Cyg**

This carbon star is identified as Gaia DR2 source ID 2062369288973870976 by SIMBAD database. The colour index Bp-Rp = 3.635 and the absolute magnitude $M_G$ = -0.88 (-0.26, +0.23) are compatible with a LPV star.

The Gaia DR2 database classifies this star as LPV candidate, with a period of 403 ± 49 days. The ASAS-SN Catalog of Variable Stars II identifies this star, ASASSN-V J201208.06+393649.6, with a different Gaia DR2 source ID 20623692889738711 and classifies this variable as Mira, with a mean magnitude V = 11.87, an amplitude of 2.38 mag and a period of 409 days.

The ASAS-SN cross-reference ID is deemed incorrect because this source cannot be found in a radius of 60 arcsec and photometric data reported in the database are those of source ID 2062369288973870976.

Our period analysis, based on 656 ASAS-SN observations in the filter V on a time span of 1377 days, applying the ANOVA method, highlighted a period of 410 ± 8 days (Figure 20), with a mean fit curve amplitude of 2.1 mag and a maximum at epoch 2458367 ± 4 HJD. The original Mira type classification and period are confirmed.

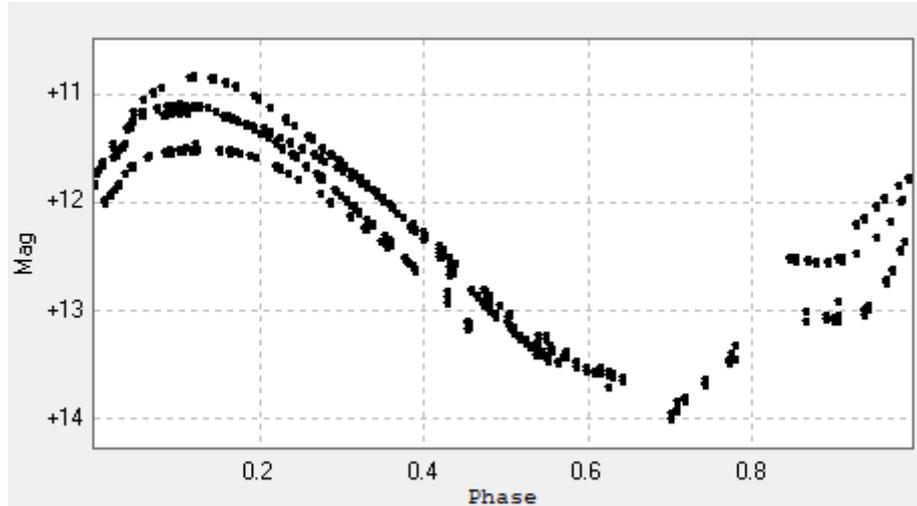

Figure 20    V1633 Cyg - V mag vs Phase (410 ± 8 days period - ANOVA)

**V1634 Cyg**

This star is identified as Gaia DR2 source ID 2074717423030592512 by SIMBAD database. The colour index Bp-Rp = 2.594 and absolute magnitude $M_G$ = 0.38 (-0.79, +0.81) are compatible with a LPV star.

The ASAS-SN Catalog of Variable Stars II reports this variable as YSO, with a mean magnitude V = 14.45 and an amplitude of 0.08 mag. This catalogue identifies this star, ASASSN-V J201219.38+415149.7, with a different Gaia DR2 source ID 2074717423024363520, with magnitude G = 14.331 and colour index Bp-Rp = 1.751.

The ASAS-SN cross-reference ID is deemed incorrect because in accordance with Gaia DR2 database such photometric characteristics apply to Gaia DR2 source ID 2074717423024363392. We performed a period analysis with 666 valid observations available from ASAS-SN in the V filter and covering a time span of 1358 days. Applying the Lomb-Scargle algorithm, we found a period of 254 ± 18 days (Figure 21), with a mean fit curve amplitude of 0.04 mag and a maximum at 2457193 ± 15 HJD.

The original Mira type classification and period are not confirmed by ASAS-SN analysis and our assessment.

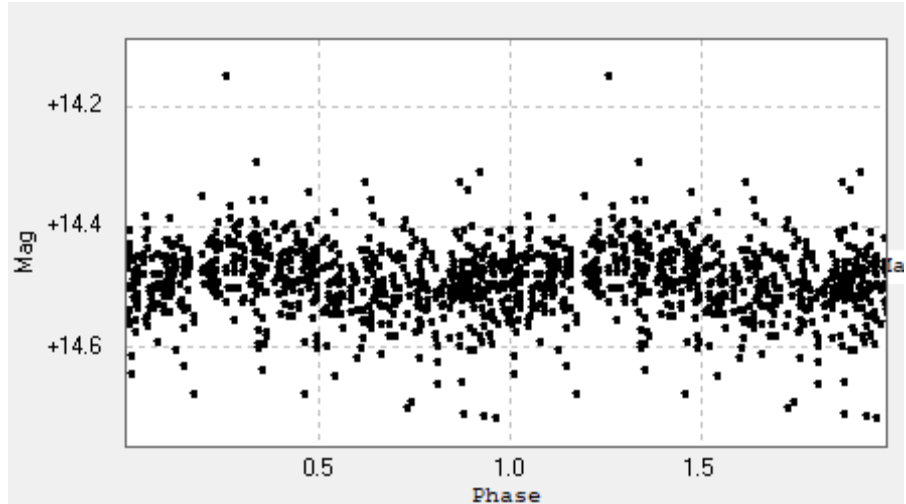

Figure 21    V1634 Cyg - V mag vs Phase (254 ± 18 days period - Lomb-Scargle)

**V1635 Cyg**

This star is identified as Gaia DR2 source ID 2061782665176034816 by SIMBAD database.

The colour index Bp-Rp = 5.684 and the absolute magnitude $M_G$ = 1.57 (-0.98, +0.86) are compatible with a LPV star.

The Gaia DR2 database classifies this star as LPV candidate, with a period of 399 ± 40 days.

The ASAS-SN Catalog of Variable Stars II identifies this star, ASASSN-V J201234.20+390233.8, with a different Gaia DR2 source ID 2061782665176035584, that is a source 10 arcsec apart from 2061782665176034816. The analysis performed by ASAS-SN classifies this variable as YSO, with a mean magnitude V = 16.09 and an amplitude of 0.36 mag.

Based on original infrared magnitude range, from equation [1] and [2], the expected mean magnitude of the variable is G ~14.8 and V ~ 15.5 and the most likely candidate that matches these values is the source ID 2061782665176034816. However, because the difference of magnitude between the two sources is ~1.1, the source ID 2061782665176035584 cannot be fully excluded. We noted that using as input the equatorial coordinates of both sources, available from Gaia DR2 database, the ASAS-SN database returns the same light curve; this seems to exclude that two close variable stars are present in a circle of 10 arcsec.

Our period analysis, based on 148 ASAS-SN observations in the filter V on a time span of 1352 days, applying the Lomb-Scargle did not highlight a reliable periodicity for this variable.

The original Mira type classification and period are confirmed by Gaia DR2 analysis but not by ASAS-SN and our assessment.

**V1636 Cyg**

This star is classified as Gaia DR2 source ID 2074640560281165184 by SIMBAD database.

It is 5821 (-1828, +2758) pc away and has a temperature of 3280 K (-4, +8).

The colour index Bp-Rp = 6.433 and absolute magnitude $M_G$ = 0.80 (-0.84, +0.82) are compatible with a LPV star. The Gaia DR2 database classifies this star as LPV candidate, with a period of 297 ± 20 days.

The ASAS-SN Catalog of Variable Stars II identifies this star, ASASSN-V J201243.76+413602.8, with a different Gaia DR2 source ID 2074640560281165312 and classifies it as Semiregular, with a mean magnitude V = 15.62, an amplitude of 0.61 mag and a period of 326 days. The ASAS-SN cross-reference ID is deemed incorrect because this source cannot be found in a radius of 60 arcsec and photometric data reported in the database are those of source ID 2074640560281165184.

In our period analysis, performed using 607 valid observations available from ASAS-SN in the V filter and covering a time span of 1358 days, applying the ANOVA algorithm, we found a period of 320 ± 11 days (Figure 22) and a maximum at epoch 2457487 ± 2 HJD.

The original Mira classification and period are confirmed and refined by Gaia DR2 and our analysis whilst Semiregular type is proposed by ASAS-SN.

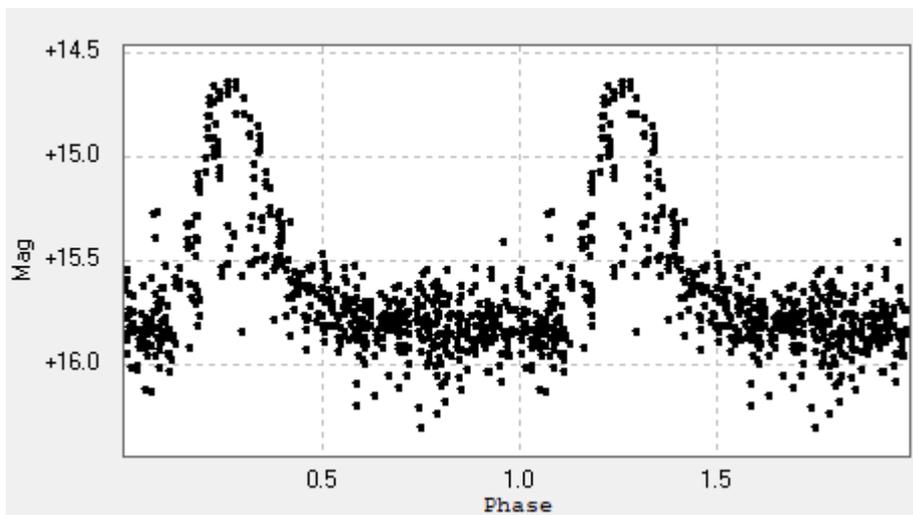

Figure 22    V1636 Cyg - V mag vs Phase (320 ± 11 days period - ANOVA)

**V1637 Cyg**

This star is identified with Gaia DR2 source ID 2062602557236419456 by SIMBAD database. The Gaia DR2 database classifies this star as LPV candidate, with a period of 212 ± 14 days. This classification is also confirmed by the position of the star in the CAMD of Figure 1, being its colour index Bp-Rp = 4.143 and absolute magnitude $M_G$ = 0.86 (-0.71, +0.66).

The ASAS-SN Catalog of Variable Stars II identifies this star, ASASSN-V J201252.20+410306.7, with a different Gaia DR2 source ID 2062602557236419584 and classifies it as Semiregular, with a mean magnitude V = 16.65, an amplitude of 1.76 mag and a period of 424 days.

The ASAS-SN cross-reference ID is deemed incorrect because this source cannot be found in a radius of 60 arcsec and photometric data reported in the database are those of source ID 2062602557236419456.

We performed a period analysis using 113 valid observations available from ASAS-SN in the V filter and covering a time span of 1101 days. Applying the Lomb-Scargle algorithm, we found a period of 267 ± 30 days (Figure 23) and a maximum at the epoch 2457660 ± 5 HJD. The original Semiregular type is confirmed with a potential revision of the period.

Our analysis with all three methods, did not find any valid period in the range of values suggested by the original and ASAS-SN analyses. As shown in Figure 24, solutions with periods around 424 days have low values of the parameter theta and are therefore very improbable.

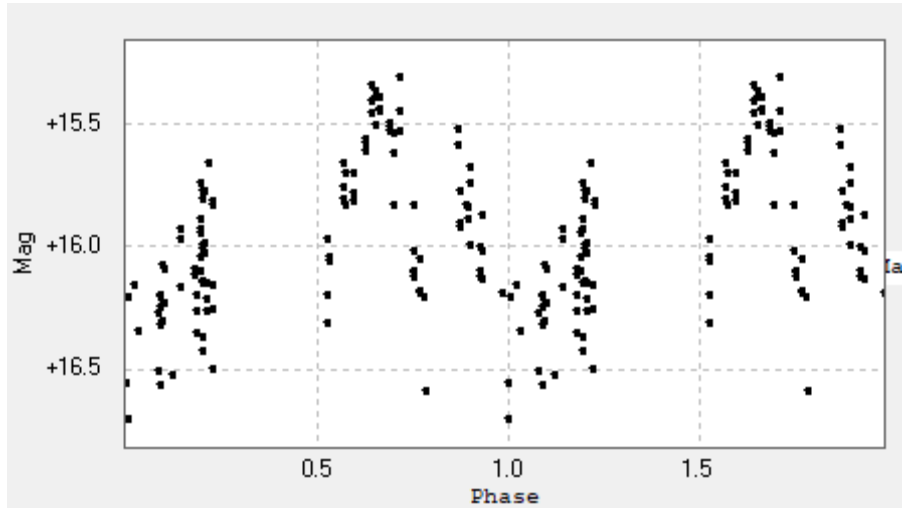

Figure 23   V1637 Cyg - V mag vs Phase (267 ± 30 days period - Lomb-Scargle)

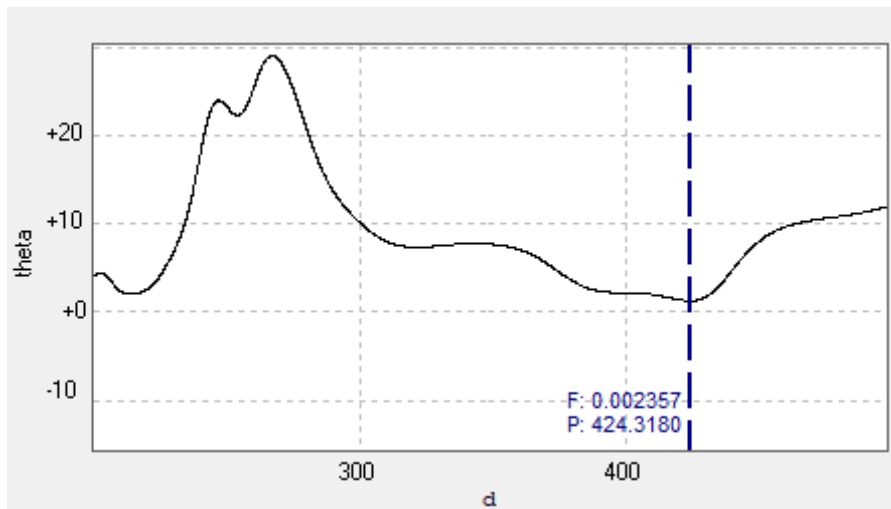

Figure 24   V1637 Cyg - Period analysis in the range 200 ÷ 500 days (Lomb-Scargle)

**V1638 Cyg**

Table 6 lists a group of six stars for which no data were available from bibliographic references reported by SIMBAD database. Also, not sufficient valid photometric measurements were available from AAVSO and ASAS-SN databases and therefore no light curve analysis was possible.

Only a general consideration on the type of variable is possible based on their photometric characteristics. For these variables, we could only conclude that their colour index Bp-Rp and absolute magnitude $M_G$ are compatible with the LPV group defined in Figure 1, and therefore with the original SR or M classification.

Table 6 summarizes photometric characteristics, the original type and period.

Table 6    Compatible LPV without any light curve analysis

| Name | Bp-Rp | M$_G$ | Original Type | Original Period (days) |
|---|---|---|---|---|
| V1638 Cyg | 6.047 | 2.04 | SR | 416 |
| V1651 Cyg | 6.663 | 3.45 | M | 383 |
| V1655 Cyg | 7.383 | 2.79 | M | 462 |
| V1660 Cyg | 6.236 | 3.84 | M | 420 |
| V1682 Cyg | 6.672 | 3.71 | SR | 290:: |
| V1694 Cyg | 6.627 | 3.60 | SR or M | 308:: |

**V1639 Cyg**

This star is identified as Gaia DR2 source ID 2061718996577182848 by SIMBAD database.

The colour index Bp-Rp = 6.266 and the absolute magnitude M$_G$ = 2.66 (-0.99, +075) are compatible with a LPV star.

The Gaia DR2 database classifies this star as LPV candidate, with a period of 371 ± 34 days. The ASAS-SN Catalog of Variable Stars II identifies this star, ASASSN-V J201323.50+383846.7, with a different Gaia DR2 source ID 2061718996577182720 and classifies this variable as Semiregular, with a mean magnitude V = 15.28, an amplitude of 0.25 mag and a period of 182 days, that is half of the original 360 days period.

The ASAS-SN cross-reference ID is deemed incorrect because this source cannot be found in a radius of 60 arcsec and photometric data reported in the database are those of source ID 2061718996577182848.

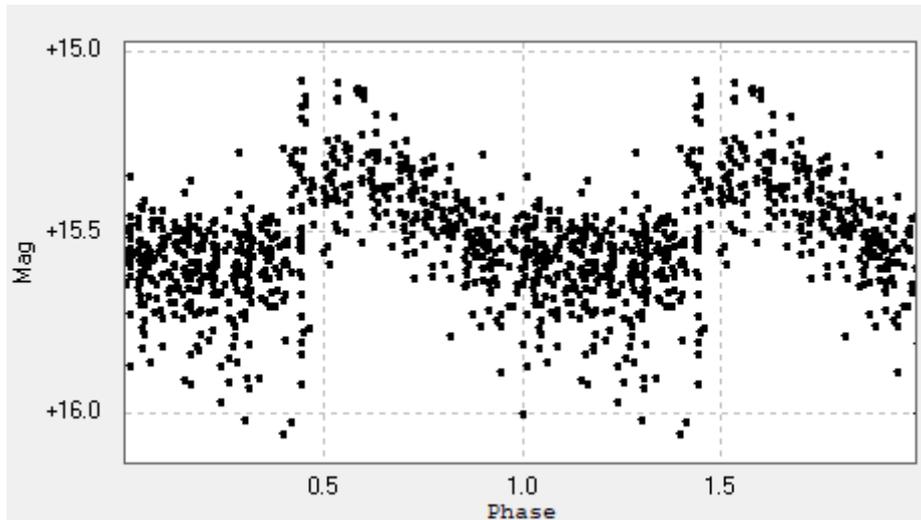

Figure 25    V1639 Cyg - V mag vs Phase (182 ± 3 days period - Lomb-Scargle)

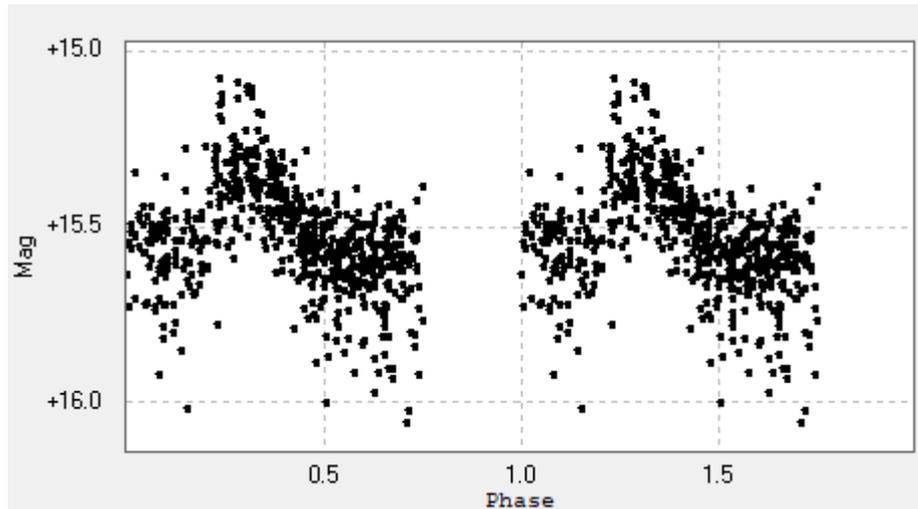

Figure 26    V1639 Cyg - V mag vs Phase (363 ± 10 days period - Lomb-Scargle)

Our period analysis, based 609 ASAS-SN observations in the filter V on a time span of 1365 days, highlighted two potential periods in accordance with ASAS-SN and Gaia DR2 results, with the dominant one depending on the applied method. The weighted average of the periods we found applying the Lomb-Scargle, ANOVA and FALC methods is: 182 ± 3 days (Figure 25) and 363 ± 10 days (Figure 26). The mean fit curve amplitude is 0.25 mag, and a maximum was identified at epoch 2457929 ± 4 HJD.

The original type and period are not confirmed by ASAS-SN analysis whilst are consistent with Gaia DR2 and one of our solutions.

**V1640 Cyg**

The colour index Bp-Rp = 6.091 and the absolute magnitude $M_G$ = 2.35 (-0.99, +0.86) are compatible with a LPV star.

The ASAS-SN Catalog of Variable Stars II classifies this variable as Semiregular, with a mean magnitude V = 16.03, an amplitude of 0.44 mag and a period of 45.96 days.

Our period analysis, based on 502 ASAS-SN observations in the filter V on a time span of 1362 days, applying the ANOVA method, highlighted two potential periods.

We found a first valid solution with a period of 29.5 ± 0.13 days, with a mean fit curve amplitude of 0.13 mag (Figure 27), and a second period at 59.0 ± 0.44 days, with a smaller mean amplitude of 0.09 mag (Figure 28).

Our analysis also identified a solution with a period of 44 days, in accordance with ASAS-SN result, but this period did not pass the period significance test we performed.

No reliable value for a maximum was identified.

The original Mira type classification and period are not confirmed.

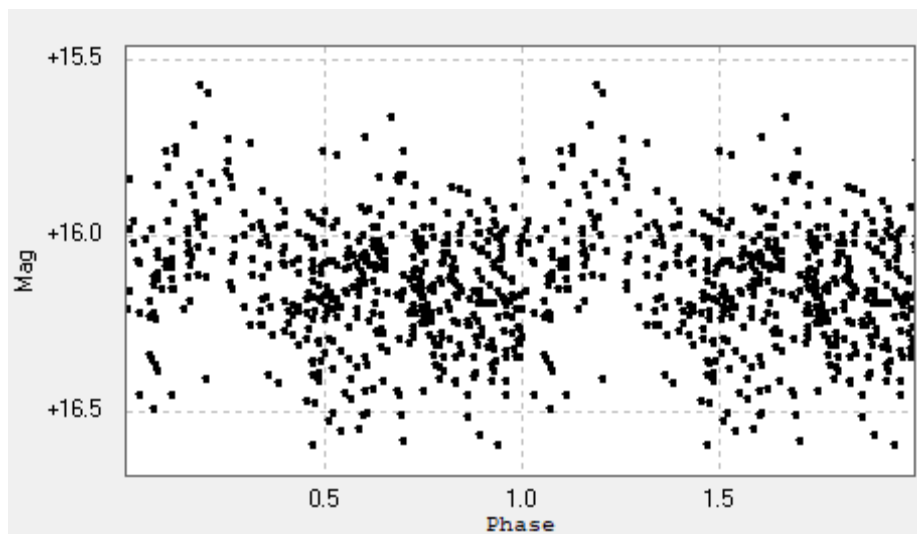

Figure 27   V1640 Cyg - V mag vs Phase (29.5 ± 0.13 days period - ANOVA)

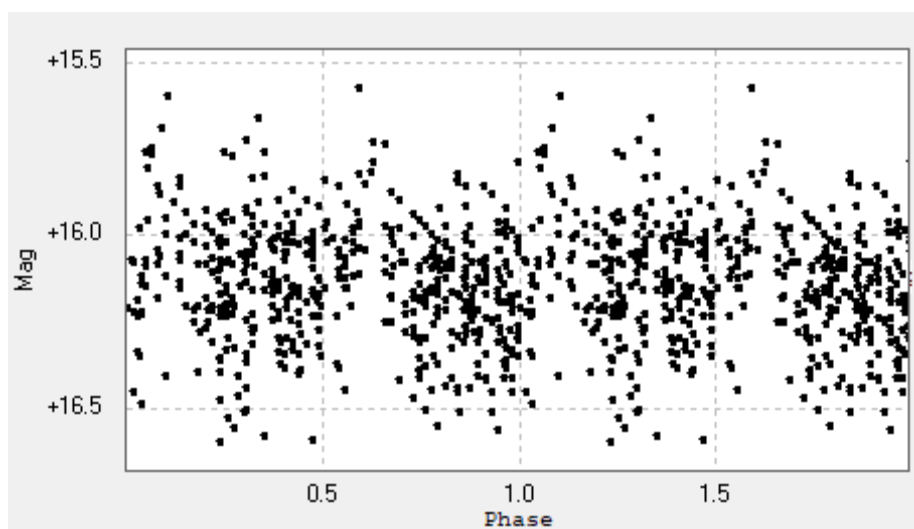

Figure 28   V1640 Cyg - V mag vs Phase (59.0 ± 0.44 days period - ANOVA)

**V1641 Cyg**

This star is identified as Gaia DR2 source ID 2061712193348895872 by SIMBAD database. Because its absolute magnitude cannot be calculated, it was not possible to compare its photometric characteristics to the CAMD of Figure 1.

The ASAS-SN Catalog of Variable Stars II identifies this star, ASASSN-V J201348.91+383807.7, with a different Gaia DR2 source ID 2061712193348896000 and classifies it as YSO, with a mean magnitude V = 16.06, and an amplitude of 0.45 mag. These two sources are separated by only 3.6 arcsec and have magnitude G = 16.208 and 16.734 respectively, that are fainter than expected from observed ASAS-SN V mean magnitude.

Moreover, images centred on equatorial coordinates of this variable does not show a couple of $16^{th}$-$17^{th}$ magnitude stars so close.

Therefore, even if the difference may be due to the invalidity of equations [1] and [2] or overestimation of original magnitude, it is probable that this star has been misidentified by both SIMBAD and ASAS-SN database. We could not identify an alternative variable source without access to the original plates (Maffei, 1977), that would allow to clarify this mismatch.

We performed a period analysis of data associated to Gaia DR2 2061712193348896000, based on 457 ASAS-SN observations in the filter V on a time span of 1362 days. Applying the Lomb-Scargle method, we highlighted a period of 525 ± 130 days (Figure 29), with a mean fit curve amplitude of 0.1 mag. No reliable value for a maximum was identified.

The original SR or M classification and period cannot be confirmed, and a misidentification of the variable is probable.

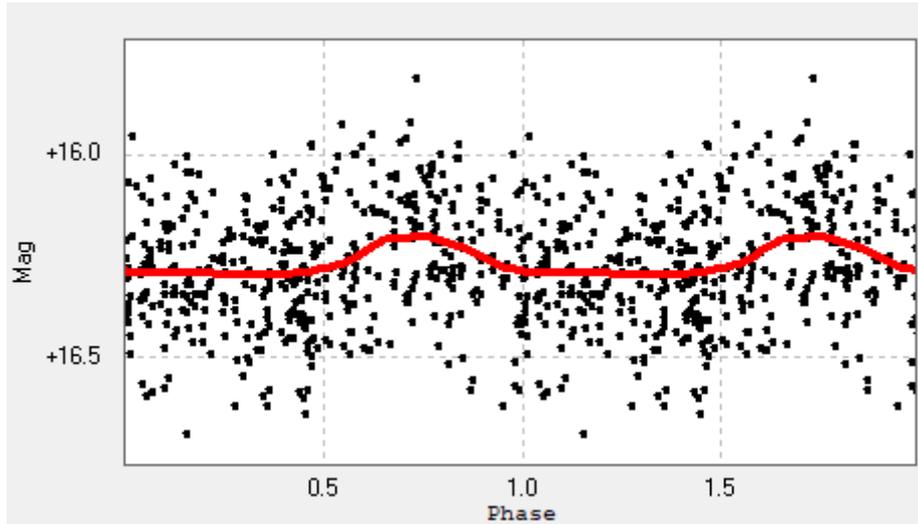

Figure 29     Gaia DR2 2061712193348896000 - V mag vs Phase (525 ± 130 days period)

**V1642 Cyg**

The colour index Bp-Rp = 5.094 and the absolute magnitude $M_G$ = 1.55 (-0.77, +0.74) are compatible with a LPV star. The Gaia DR2 database classifies this star as LPV candidate, with a period of 296 ± 24 days.

The ASAS-SN Catalog of Variable Stars II classifies this variable as Semiregular, with a mean magnitude V = 17.25, an amplitude of 0.6 mag and a period of 7.2068939 days. We did not perform any analysis of the light curve because not sufficient valid observations were available. The ASAS-SN results are not consistent with the result of the Gaia DR2 analysis and the original Mira type classification and period of 290 days.

**V1643 Cyg**

This star is identified with Gaia DR2 source ID 2060950884625235584 by SIMBAD database. The colour index Bp-Rp = 6.233 and the absolute magnitude $M_G$ = 1.95 (-1.00, +0.82) are compatible with a LPV star.

The Gaia DR2 database classifies this star as LPV candidate, with a period of 305 ± 23 days. The ASAS-SN Catalog of Variable Stars II identifies this star, ASASSN-V J201420.07+382745.7, with a different Gaia DR2 source ID 2060950884625235712 and classifies it as Semiregular, with a mean magnitude V = 12.49, an amplitude of 0.07 mag and a period of 289 days.

The ASAS-SN cross-reference ID is deemed incorrect because refers to a source of magnitude G = 17.1537, that is inconsistent with the observed mean V mag and the original infrared range.
Our period analysis, based on 655 ASAS-SN observations in the filter V on a time span of 1377 days, applying the Lomb-Scargle method highlighted a period of 306 ± 29 days (Figure 30), with a mean fit curve amplitude of 0.08 mag and a maximum at epoch 2457598 ± 4 HJD.
The original Mira type classification is not fully confirmed by all analyses, but the period is confirmed.

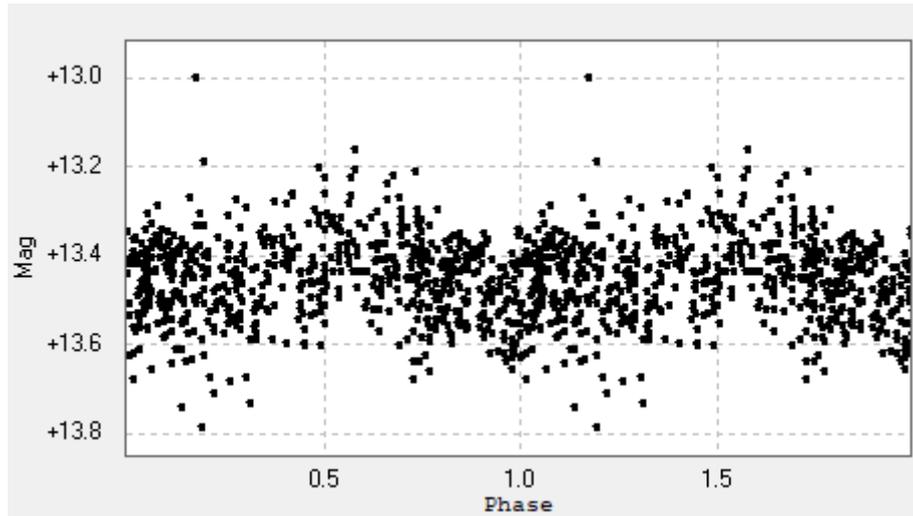

Figure 30    V1643 Cyg - V mag vs Phase (306 ± 29 days period - Lomb-Scargle)

**V1645 Cyg**
The colour index Bp-Rp = 6.175 and the absolute magnitude $M_G$ = 1.97 (-0.91, +0.79) are compatible with a LPV star.
The Gaia DR2 database classifies this star as LPV candidate, with a period of 354 ± 34 days.
We did not perform any period analysis because there were not sufficient valid data in the ASAS-SN database.
The original Mira classification and period are confirmed by Gaia DR2 light curve analysis.

**V1646 Cyg**
A group of four stars, whose colour index Bp-Rp is not available is listed in Table 7.
No data were available from bibliographic references reported by SIMBAD database and no sufficient valid photometric measurements were available from AAVSO and/or ASAS-SN databases.
Therefore, it was not possible to analyse their light curve, determine their position in the CAMD of Figure 1 and assess the original type and period.
Table 7 highlights the missing data and summarizes the original classification.

Table 7       Variable stars with no assessment

| Name | Bp-Rp | Nominal Distance (pc) | Gaia DR2 Median G (mag.) | $M_G$ | Original Type | Original Period (days) |
|---|---|---|---|---|---|---|
| V1646 Cyg | - | 1905 | 15.755 | 4.36 | M: | 388 |
| V1657 Cyg | - | 2155 | 15.182 | 3.51 | M | 310 |
| V1688 Cyg | - | 3112 | 15.390 | 2.93 | M | 312: |
| V1689 Cyg | - | 5461 | 15.318 | 1.63 | SR or M | 484: |

**V1647 Cyg**

This star was originally classified as Semiregular or Mira type with an uncertain period of 334 days. Based on its colour index Bp-Rp = 3.797 and absolute magnitude $M_G$ = 6.80 (-0.89, +0.66) we classified this variable as YSO. This classification is consistent with our period analysis of 435 valid observations available from ASAS-SN in the V filter and covering a time span of 1357 days, that, as expected, did not highlight any periodicity for this variable.

**V1648 Cyg**

The colour index Bp-Rp = 5.199 and absolute magnitude $M_G$ = 2.67 (-0.86, +0.77) are compatible with a LPV star. The ASAS-SN photometry database classifies this variable as Semiregular, with an amplitude of 0.35 mag and a period of 286 days. Because only 37 valid observations were available from ASAS-SN database, we could not identify any reliable solution for a potential period. The original Mira type classification is not confirmed by ASAS-SN photometry, whilst the result of the period analysis is consistent with the known value of 300 days.

**V1649 Cyg**

This star is identified by SIMBAD database with Gaia DR2 source ID 2062342728892073088 and WISE J201828.98+403406.5. Its colour index Bp-Rp = 6.615 and absolute magnitude $M_G$ = 5.06 (-0.82, +0.61) are compatible with a star on the border between the Semiregular/Irregular and YSO groups of Figure 1.

The ASAS-SN Catalog of Variable Stars II reports this variable as Semiregular, with a mean magnitude V = 16.77, an amplitude of 0.44 magnitudes and a period of 226 days. It also identifies this star, ASASSN-V J201828.99+403406.3, with a different Gaia DR2 2062342728892072960 whilst the WISE identification number is in accordance with the one specified by SIMBAD. The ASAS-SN cross-reference ID is deemed incorrect because that this source cannot be found in a radius of 60 arcsec and photometric data and WISE source ID reported in the database are those of source Gaia DR2 2062342728892073088.

Based on the ASAS-SN photometric data in the V filter, we performed a period analysis using 42 valid observations covering a time span of 1325 days, but we did not find any reliable periodicity. The original classification and period are not confirmed.

**V1650 Cyg**

This star is identified as Gaia DR2 source ID 2061470128971060864 by SIMBAD database and has a spectral type M6-M7 (Stephenson, 1992).

Its colour index Bp-Rp = 6.128 and absolute magnitude $M_G$ = 0.18 (-0.89, +0.71) are compatible with a LPV star.

The Gaia DR2 database classifies this star as LPV candidate, with a period of 391 ± 58 days. The ASAS-SN Catalog of Variable Stars II identifies this star, ASASSN-V J201838.07+391756.0, with a different Gaia DR2 source ID 2061470128971060736, and classifies it as Semiregular, with a mean magnitude V = 14.49, an amplitude of 0.69 mag and a period of 388 days. The ASAS-SN cross-reference ID is deemed incorrect because that this source cannot be found in a radius of 60 arcsec and photometric data reported in the database are those of source Gaia DR2 2061470128971060864.

We performed our period analysis using 647 ASAS-SN observations covering a time span of 1377 days. Applying the Lomb-Scargle method, we found a period of 385 ± 30 days (Figure 31). Insufficient observations were available to determine the epoch of a maximum.

The original classification of V1650 Cyg is confirmed and the period is refined.

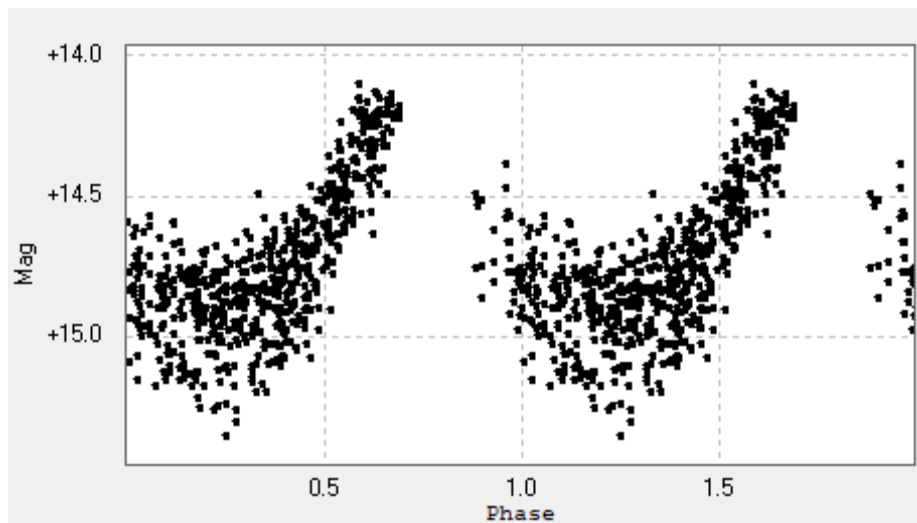

Figure 31    V1650 Cyg - V mag vs Phase (385 ± 30 days period - Lomb-Scargle)

**V1651 Cyg**

See the assessment of V1638 Cyg.

**V1652 Cyg**

The colour index Bp-Rp = 6.063 and absolute magnitude $M_G$ = 5.93 (-0.44, +0.37) are compatible with a variable at the border between the Semiregular and YSO groups of Figure 1. We did not perform period analysis for this star because not sufficient valid photometric data were available.

The original Mira classification and uncertain period of 565 days are not confirmed.

**V1653 Cyg**

This star is identified as Gaia DR2 source ID 2061416012387664512 by SIMBAD database.
The Gaia DR2 database classifies this star as LPV candidate, with a period of 314 ± 79 days. This classification is confirmed also by its colour index Bp-Rp = 6.420 and absolute magnitude $M_G$ = -0.28 (-0.74, +0.80), which place this star in the LPV group of Figure 1.

The ASAS-SN Catalog of Variable Stars II identifies this star, ASASSN-V J202043.11+392239.9, with a different Gaia DR2 source ID 2061416012387664384 and classifies it as Semiregular, with a mean magnitude V = 15.03, an amplitude of 0.29 mag and a period of 7.06814 days. The ASAS-SN cross-reference ID is deemed incorrect because that this source cannot be found in a radius of 60 arcsec and photometric data reported in the database are those of source Gaia DR2 2061416012387664512. Based on ASAS-SN photometric data in the V filter, our period analysis with Lomb-Scargle method of 113 observations, on a time span of 1101 days, highlighted a period of 267 ± 31 days (see Figure 32), with a maximum at epoch 2457660 ± 5 HJD. Our analysis also identified a solution with a period of 7.58 days, in accordance with ASAS-SN result, but this period did not pass the period significance test we performed.

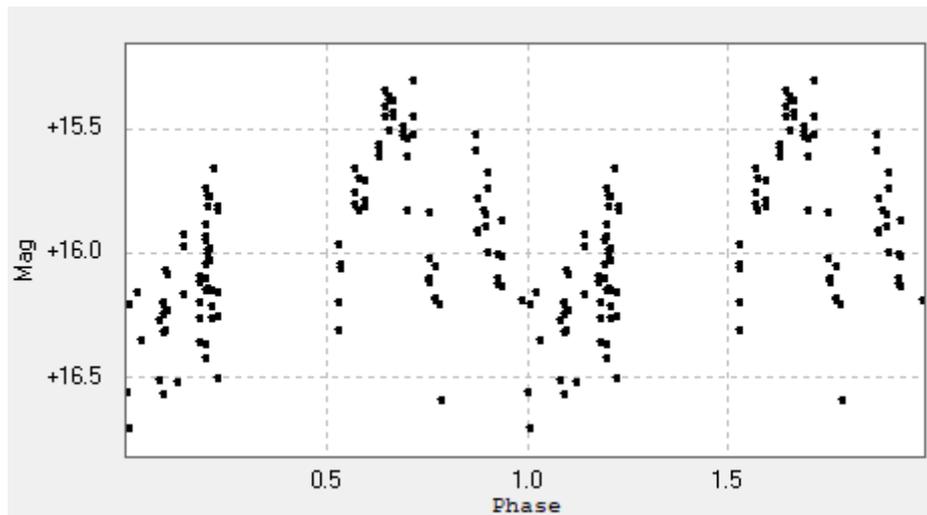

Figure 32    V1653 Cyg – V mag vs Phase (267 ± 31 days period - Lomb-Scargle)

**V1654 Cyg**

This star is 5304 (-1465, +2091) pc away and has a temperature of 3281 ± 6 K.
The colour index Bp-Rp = 6.010 and absolute magnitude $M_G$ = 0.46 (-0.72, +0.70) are compatible with a LPV star. The ASAS-SN Catalog of Variable Stars II reports this variable as Semiregular, with a mean magnitude V = 12.45, an amplitude of 0.07 magnitudes and a period of 252 days.

We noted that the amplitude found by ASAS-SN analysis is not consistent with the original amplitude of 4.3 mag. measured using a photographic infrared filter. We performed a period analysis using 674 valid observations available from ASAS-SN in the V filter and covering a time span of 1358 days. Applying the Lomb-Scargle method, we found a period of 262 ± 19 days (Figure 33), an amplitude of 0.2 mag and a maximum at epoch 2458362 ± 2 HJD.

The original Mira classification is not confirmed whilst values found for the period are consistent with the original one.

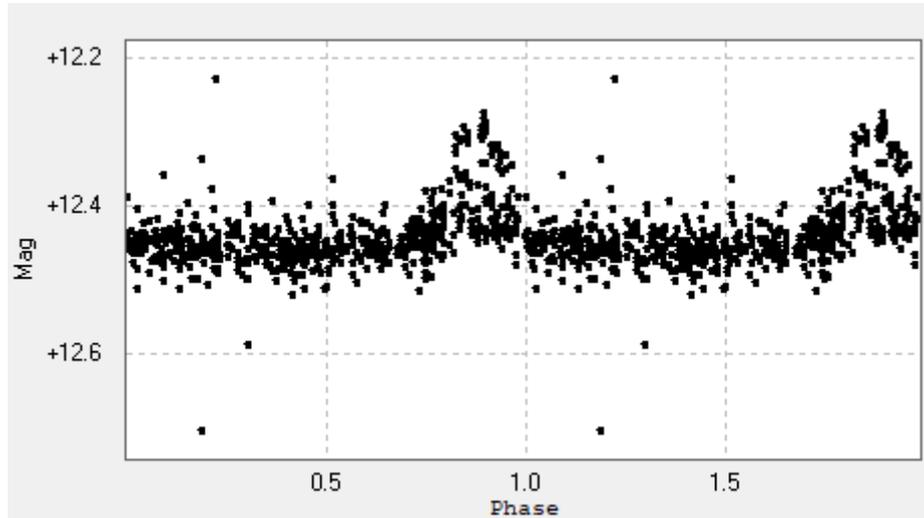

Figure 33    V1654 Cyg – V mag vs Phase (262 ± 19 days period - Lomb-Scargle)

**V1655 Cyg**

See the assessment of V1638 Cyg.

**V1656 Cyg**

This star is 3916 (-1124, +1760) pc away and has a temperature of 3281 (-6, +8) K. The IBVS 3800 (Zsoldos, 1991) classifies this star as M7 or M9.

The colour index Bp-Rp = 6.184 and the absolute magnitude $M_G$ = 0.62 (-0.81, +0.73) are compatible with a LPV star.

The Gaia DR2 database classifies this star as LPV candidate, with a period of 314 ± 34 days.

The ASAS-SN Catalog of Variable Stars II reports this variable as Semiregular, with a mean magnitude V = 16.45, an amplitude of 0.7 mag and a period of 291 days.

We performed a period analysis using 175 valid observations available from ASAS-SN in the V filter and covering a time span of 1296 days. Applying the Lomb-Scargle method we found a dominant period of 148 ± 9 days (Figure 34) and a second prominent period of 295 ± 28 days (Figure 35). Vice versa, when performing the analysis using both ANOVA and FALC method we found a dominant period of 295 ± 15 days. The mean amplitude of the curve fit is in the range from 0.38, for the shorter period, to 0.40 mag, for the longer one, in the V filter. A maximum of the light curve was identified at epoch 2457593 ± 7 HJD.

The shorter period is approximately half of the longer one but, due to its high significance, is not considered an alias and is a candidate for more detailed long-term photometric observations. The Mira type of V1656 Cyg is not confirmed by ASAS-SN light curve, which classifies this star as Semiregular. Our analysis confirms the compatibility of the photometric characteristics with those of Mira type variable and suggests a shorter period of 148 days.

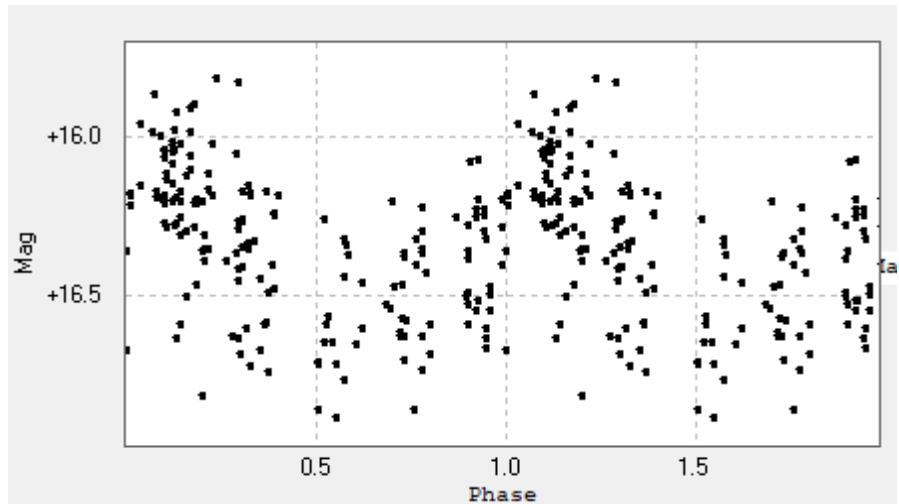

Figure 34    V1656 Cyg - V mag vs Phase (148 ± 9 days period - Lomb-Scargle)

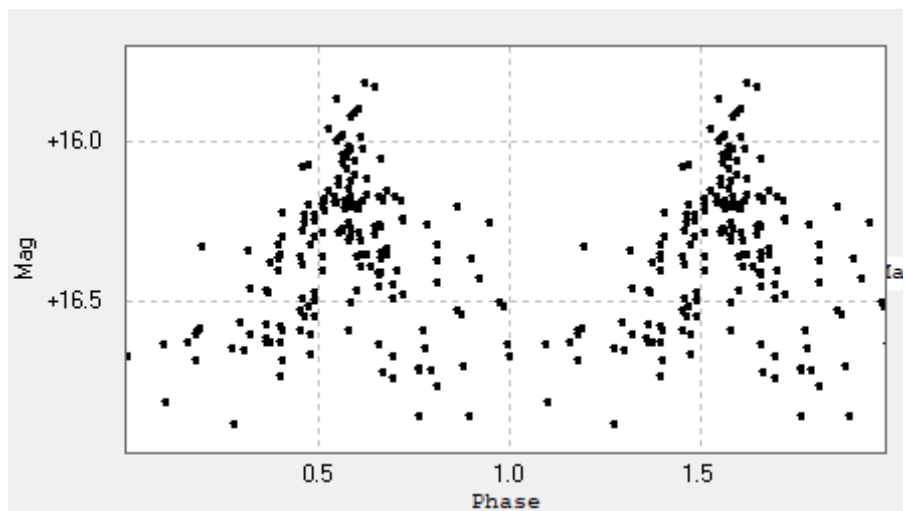

Figure 35    V1656 Cyg - V mag vs Phase (295 ± 28 days period - ANOVA/FALC)

**V1657 Cyg**
See assessment of V1646 Cyg.

**V1658 Cyg**
This is an extremely red star with a colour index, with an effective temperature of 3281 (-5, +7) K and a spectral type M9 (Stephenson, 1992). The colour index Bp-Rp = 7.133 and the absolute magnitude $M_G$ = 1.63 (-0.87, +0.70) are compatible with a LPV star.

The Gaia DR2 database classifies this star as LPV candidate, with a period of 413 ± 39 days. The ASAS-SN Catalog of Variable Stars II classifies this variable as Mira, with a mean magnitude V = 16.75, an amplitude of 2.06 mag and a period of 196 days.

We performed a period analysis using 73 valid observations available from ASAS-SN in the V filter and covering a time span of 1205 days. Applying the Lomb-Scargle, ANOVA and FALC algorithms, we found a dominant period of 112 ± 1 days (Figure 36), with a mean amplitude of the curve fit of 0.72 mag in the V filter and a maximum at epoch 2457895 ± 2 HJD.

No significative period was found around the values of 196, 382 or 413 days found in the original work, ASAS-SN and Gaia DR2 analysis. It is remarked that this value is quite short and unusual for Mira type variable, suggesting a potential different Semiregular class for this variable. The Mira type of V1658 Cyg is confirmed by ASAS-SN light of curve and its physical and photometric characteristics. Our analysis shows that the current value of the period is unlikely and highlights a potential shorter value of 112 days.

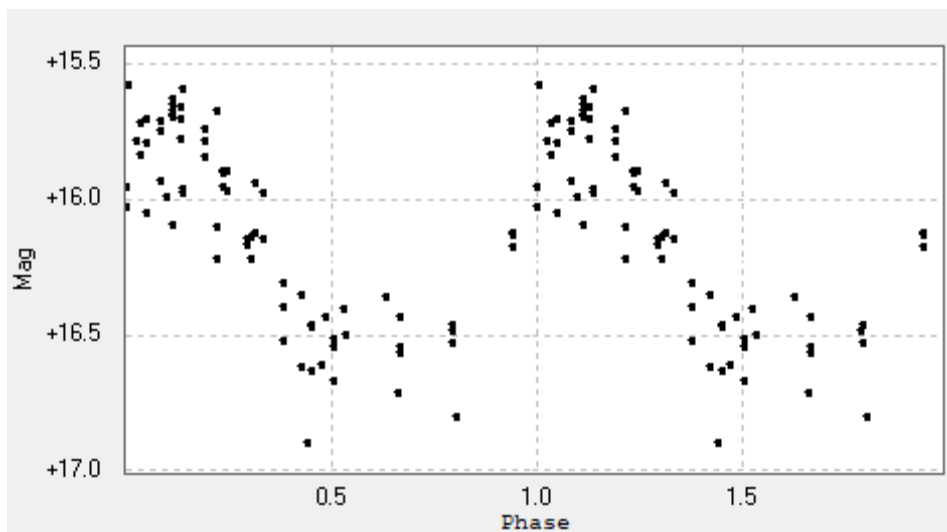

Figure 36  V1658 Cyg - V mag vs Phase (112 ± 1 days period - Lomb-Scargle)

**V1659 Cyg**

The presence of VO bands in the 'INT Photometric Hα Survey of the Northern Galactic Plane' (IPHAS) spectra of this object, as well as the TiO bands at 0.93 and 1.10 μm, marks this star as a late M6.5 giant. Based on the observed infrared excesses, it was shown that its highly reddened colours are a product of both interstellar and circumstellar reddening (Wright, 2008).

The Gaia DR2 database classifies this star as a red supergiant and LPV candidate, with a period of 491 ± 118 days. The membership to this group is also confirmed by its colour index Bp-Rp = 6.843 and absolute magnitude $M_G$ = 1.86 (-0.82, +0.79).

In the ASAS-SN photometry database there were no sufficient valid observations to build a light curve and estimate the period. We performed a period analysis using untransformed measurements available from AAVSO (Figure 37) but, due to the short time span of measurements, no reliable solutions were found. The range magnitude estimated from AAVSO observations is in the range 15.8 to 19.5 in filter V, with a maximum at epoch 2458904 ± 13 HJD.

We concluded that the original Mira type classification for V1659 Cyg is compatible with the available photometric and spectroscopic characteristics and the Gaia DR2 analysis, but the period may be shorter than the original value period of 770 days.

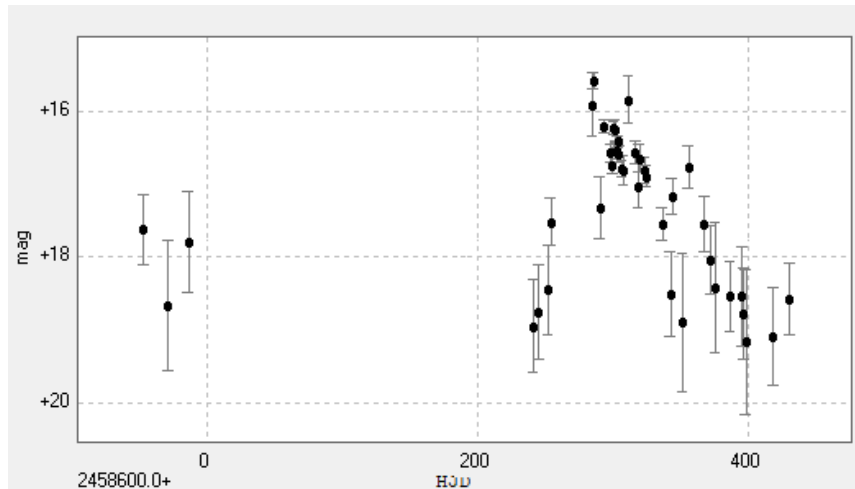

Figure 37   Untransformed AAVSO observations of V1659 Cyg

**V1660 Cyg**
See assessment of V1638 Cyg.

**V1677 Cyg**
The colour Bp-Rp = 5.572 and absolute magnitude $M_G$ = 0.16 (-0.86, +0.82) are compatible with a LPV star. The ASAS-SN Catalog of Variable Stars II reports this variable as Irregular, with a mean magnitude V = 15.07 and an amplitude of 0.15 mag. We performed a period analysis using 656 valid observations available from ASAS-SN in the V filter and covering a time span of 1358 days. Applying the Lomb-Scargle method, we found a potential period of 29.5 ± 0.2 days (Figure 38), a mean curve fit amplitude of 0.06 mag and a maximum at epoch 2457573 ± 5 HJD. We noted that the original amplitude of this variable is 0.9 mag in the photographic infrared filter. This may be due to irregular flares which add up to a small pulsation of a few magnitudes.

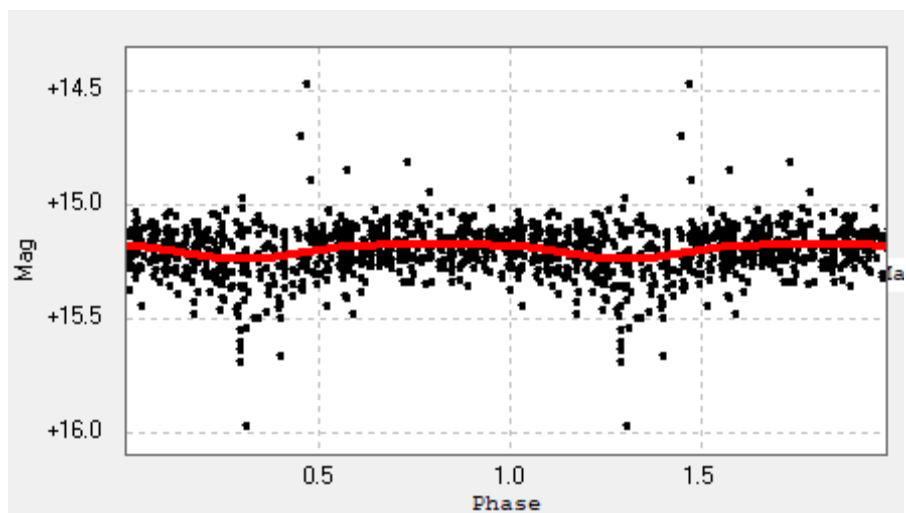

Figure 38   V1677 Cyg - V mag vs Phase (29.5 ± 0.2 days period - Lomb-Scargle)

**V1678 Cyg**

This star is identified as Gaia DR2 source ID 2061706455272623744 by SIMBAD database. Based on its colour index Bp-Rp = 0.919 and absolute magnitude $M_G$ = 2.27 (-0.47, +0.40) we excluded its membership to the LPV group.

The ASAS-SN Catalog of Variable Stars II identifies this star, ASASSN-V J201326.99+383432.2, with a different Gaia DR2 source ID 2061706455272623616 and classifies this variable as an eclipsing β Lyrae binary, with a mean magnitude V = 12.96, an amplitude of 0.41 mag and a period of 0.52784 days.

The ASAS-SN cross-reference ID is deemed incorrect because this source cannot be found in a radius of 60 arcsec and photometric data reported in the database are those of source Gaia DR2 2061706455272623744.

Our period analysis, based on 655 ASAS-SN observations in the filter V on a time span of 1377 days, applying the ANOVA method confirmed the β Lyrae type with a period of 0.5278 ± 0.0002 days that confirms the ASAS-SN result (Figure 39). The original eclipse binary classification is confirmed.

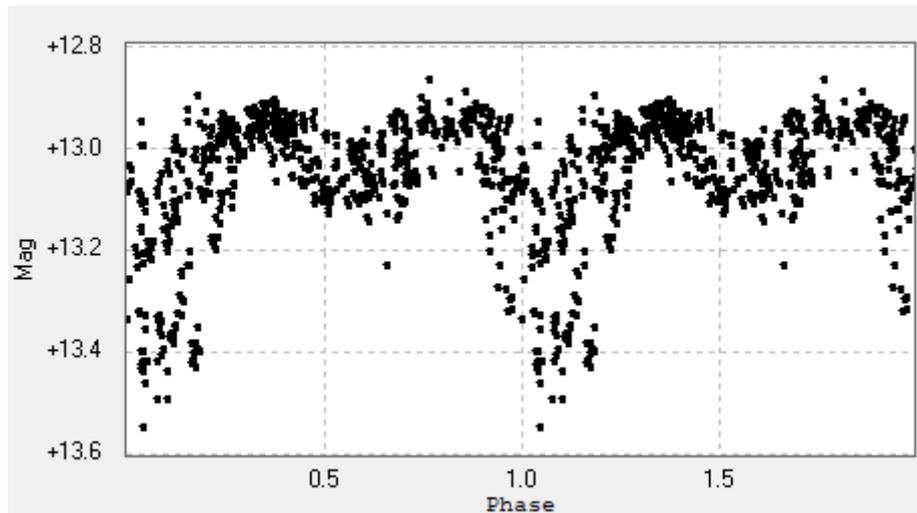

Figure 39    V1678 Cyg - V mag vs Phase (0.5278 ± 0.0002 days period - ANOVA)

**V1680 Cyg**

This star is classified as Semiregular, without a defined period, in the original work. Based on Gaia DR2 data, the distance of this star is 2932 (-1014, +1988) pc, with an effective temperature of 3284 (-9, +42) K, typical of a late spectral type. Its colour index Bp-Rp = 5.646 and absolute magnitude $M_G$ = 4.70 (-1.12, +0.92) locate this star on the border between the Semiregular/Irregular and the YSO groups of CAMD in Figure 1.

We performed an analysis of the period, using 677 valid measurements in the V filter available from ASAS-SN database, covering a time span of 1358 days. Applying the FALC method, we highlighted a dominant period of 215 ± 11 days (see Figure 40), with a mean fit curve amplitude of 0.12 mag.

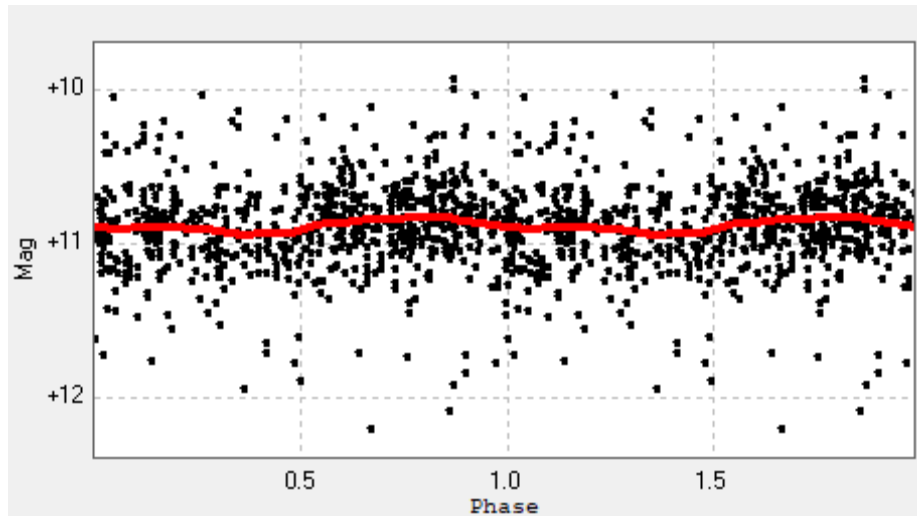

Figure 40    V1680 Cyg - V mag vs Phase (215 ± 11 days period - FALC)

**V1681 Cyg**

This star is 4883 (-1537, +2377) pc away and has a temperature of 3325 (-46, +87) K.
Its colour index Bp-Rp = 5.062 and absolute magnitude $M_G$ = 3.28 (-0.86, +0.82) are compatible with a LPV star. We performed an analysis of the period, using 269 valid measurements in the V filter of ASAS-SN, covering a time span of 1330 days. Applying the Lomb-Scargle and FALC methods we found two potential period at 27 ± 0.2 days (Figure 41) and 55 ± 0.4 days (Figure 42), with a mean amplitude for the fit curve of 0.2 mag. We did not find any reliable solution around the original, uncertain, period of 280 days. The original Semiregular type is compatible with its photometric characteristics but there is a strong evidence, from our analysis, that the period has to be revised to a shorter value.

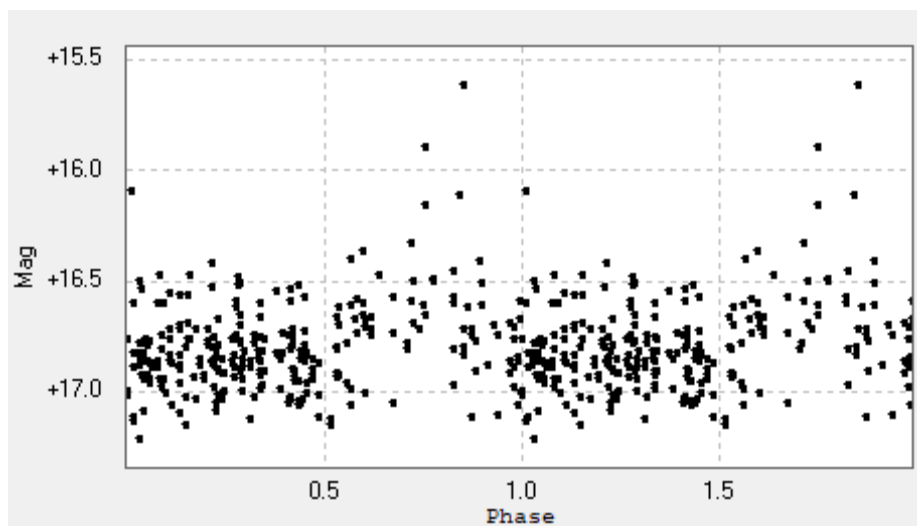

Figure 41    V1681 Cyg - V mag vs Phase (27 ± 0.2 days period - Lomb-Scargle)

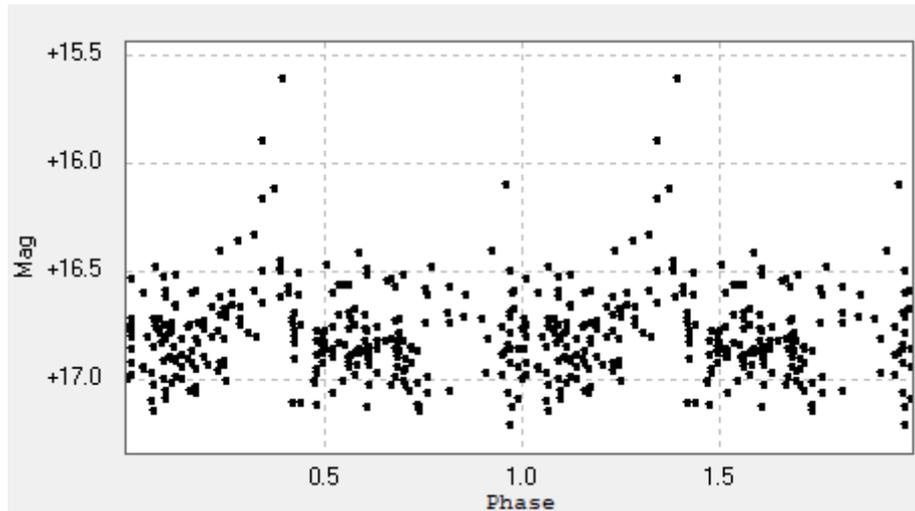

Figure 42    V1681 Cyg - V mag vs Phase (55 ± 0.4 days period - FALC)

**V1682 Cyg**
See assessment of V1638 Cyg.

**V1683 Cyg**
The colour index Bp-Rp = 1.834 and the absolute magnitude $M_G$ = 5.28 ± 0.06 are not compatible with a LPV star.
The ASAS-SN Catalog of Variable Stars II classifies this star as a generic variable, with a mean magnitude V = 16.13 and an amplitude of 1.57 mag.
Our period analysis, based on 655 ASAS-SN observations in the filter V on a time span of 1377 days, applying ANOVA method did not identify any reliable periodicity.
The original irregularity of this variable is confirmed.

**V1684 Cyg**
This star is identified as Gaia DR2 source ID 2068820948319278976 by SIMBAD database.
It is 6237 (-1584, +2303) pc away, has a temperature of 3419 (-126, +211) K and is classified as a Carbon star in the General Catalog of galactic Carbon stars (Alksnis et al, 2001).
The colour index Bp-Rp = 4.353 and the absolute magnitude $M_G$ = -0.44 (-0.68, +0.64) are compatible with a LPV star.
The Gaia DR2 database classifies this star as a red supergiant and LPV candidate, with a period of 540 ± 194 days.
The ASAS-SN Catalog of Variable Stars II identifies this star, ASASSN-V J202003.05+430257.6, with a different Gaia DR2 source ID 2068820948319279104 and reports this variable as Semiregular, with a mean magnitude V = 16.42, an amplitude of 1.89 mag and a period of 395 days.
The ASAS-SN cross-reference ID is deemed incorrect because this source cannot be found in a radius of 60 arcsec and photometric data reported in the database are those of source Gaia DR2 2068820948319278976.

We performed a period analysis using 283 valid observations available from ASAS-SN in the V filter and covering a time span of 1358 days, finding two potential solutions: 210 ± 14 days with the Lomb-Scargle algorithm (Figure 43) and 424 ± 25 days applying the ANOVA method (Figure 44). A maximum was found at epoch 2458303 ± 10 HJD.

The original Semiregular type classification is confirmed by ASAS-SN observations and photometric characteristics. Based on our analysis, the original uncertain period of 400 is likely to be shorter.

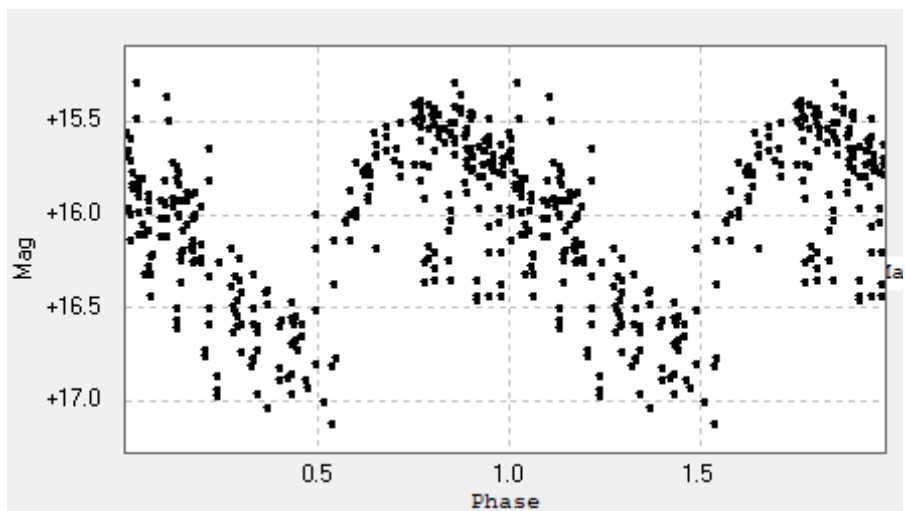

Figure 43    V1684 Cyg – V mag vs Phase (210 ± 14 days period - Lomb Scargle)

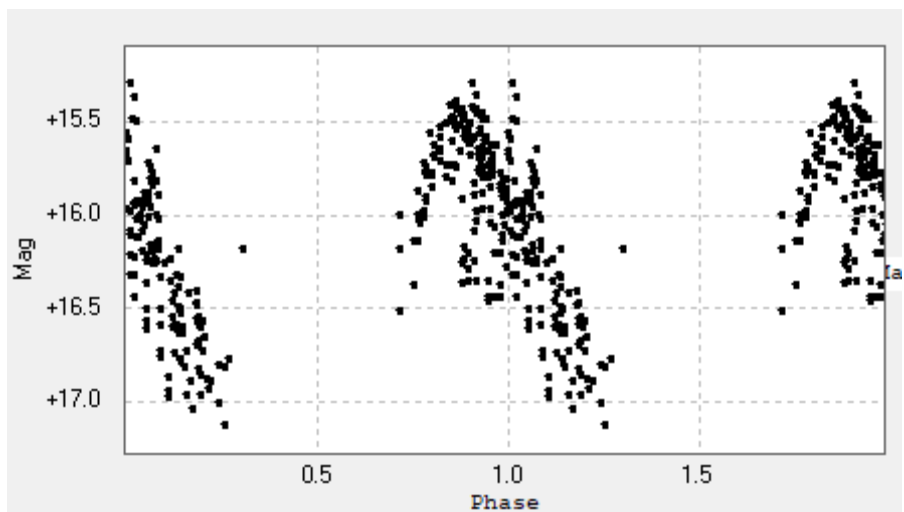

Figure 44    V1684 Cyg – V mag vs Phase (424 ± 25 days period - ANOVA)

**V1688 Cyg**
See assessment of V1646 Cyg.

**V1689 Cyg**
See assessment of V1646 Cyg.

**V1690 Cyg**

For this star, the SIMBAD database does not define a Gaia DR2 identifier. However, there is only one star in an area of 5 arcsec of radius, centred at the equatorial coordinates R.A. 20h 22m 30.92856s, Decl. +42° 49' 36.5268" available from 2MASS catalogue (Cutri et al, 2003).

We could, therefore, identify the variable V1690 Cyg as the Gaia DR2 Source ID 2069127913922504704. This star is 1184 (-197, +292) pc away and has a temperature of 3284 (-8, +51) K. The colour index Bp-Rp = 6.405 and absolute magnitude $M_G$ = 3.36 (-0.48, +0.40) are compatible with a LPV star.

We performed a period analysis using 668 observations, covering a time span of 1358 days, retrieved from the ASAS-SN database. Using the Lomb-Scargle, ANOVA and FALC methods we estimated a dominant period of 140 ± 2 days (Figure 45) between two adjacent minima, with a maximum at 2457962 ± 3 HJD and a mean amplitude of the curve fit of 0.23 mag.

We noted that the value we found is approximately half of the original one of 285 days that we found as a second prominent period of 279 ± 6 days, when we performed the analysis using only the ANOVA and FALC methods.

With respect to the original RV Tauri classification, it was observed (Zsoldos, 1991) that the proposed period of 285 days does not match with the definition of this class of variable star, whose 'formal pulsation' (the period between two adjacent primary minima) is between 30 and 150 days. In this paper we also note that the original work does not specify whether the period refers to a formal pulsation or not.

The ASAS-SN observations show a quite regular light curve, so we could not identify the typical RV Tauri pulsation, with subsequent primary (deep) and secondary (shallow) minima.

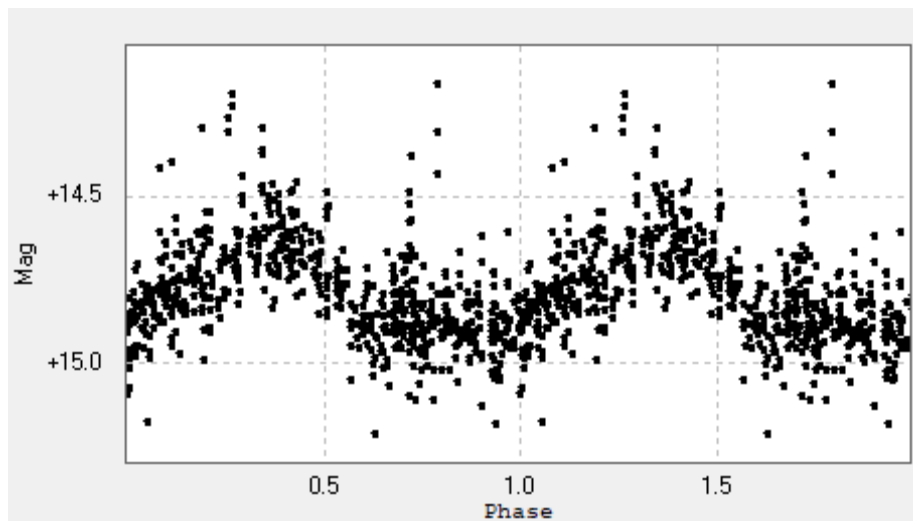

Figure 45    V1690 Cyg – V mag vs Phase (140 ± 2 days period - ANOVA)

Finally, we compared the position of the variable with respect to a WISE colour-colour diagram (Figure 46) of known galactic RV Tauri variables (Gezer et al, 2015). The colour indexes of the variable (0.44, 0.99) fall in the 'disc box' of Figure 46, which collects galactic RV Tauri with no IR excess.

For the variable V1690 Cyg there are indications that both the RV Tauri type and the period may be not correct; long-term observations are required to highlight the characteristic waveform of this class of variable, with deep minima which alternate to shallow ones and to determine the correct period. Our analysis suggests a potential shorter period of 140 days.

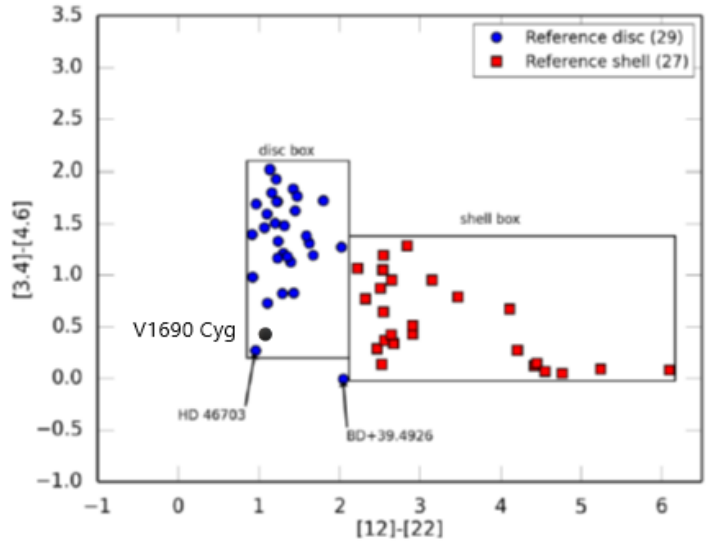

Figure 46    V1690 Cyg in the WISE colour-colour diagram for Galactic RV Tauri

**V1691 Cyg**
This 16$^{th}$ magnitude star, in the filter G, identified in the SIMBAD database as Gaia DR2 source ID 2061227136905659392, is 3071 (-927, +1548) pc away and has an absolute magnitude $M_G$ = 3.59 (-0.89, +0.78).

Because no Bp-Rp colour index is available, it is not possible to try to classify this object with respect to the groups defined in the CAMD of Figure 1.

The ASAS-SN Catalog of Variable Stars II reports this star as a generic variable, with undefined type and period and a mean magnitude amplitude of 0.29 mag. Our analysis of the ASAS-SN photometric data, using 1191 valid observations in the V filter and covering a time span of 1373 days, did not provide reliable solutions for the period.

We noted that the amplitude observed by ASAS-SN is smaller than the original 2.1 infrared magnitude range; also, the absence of any reliable solution for the period is unlikely for a candidate Mira. These inconsistencies may be due to an incorrect identification of the star. In a radius of 1' centred on the coordinates of Gaia DR2 2061227136905659392 star, R.A. 20h 23m 02.5492050797s +39° 04' 23.077997386", there are four stars with a similar magnitude, listed in Table 8. Our analysis of the ASAS-SN photometric data of these four stars did not highlight a significant variability; if a star misidentification occurred this does not involve stars within 60 arcsec from the current known position of V1691 Cyg.

Table 8        Similar sources close to V1691 Cyg

| Gaia DR2 source ID | Gaia DR2 Median G (mag.) | Distance from V1658 Cyg position (arcmin) |
|---|---|---|
| 2061227136905657088 | 14.578 | 0.2129 |
| 2061227510557912064 | 14.667 | 0.4529 |
| 2061227514862779520 | 15.976 | 0.5586 |
| 2061227205625331968 | 14.753 | 0.6749 |

**V1692 Cyg**

In accordance with the classification of IRAS low-resolution spectra (Kwok, 1997), the spectrum of this object shows a flat continuum with unusual features, whose source is unknown. No valid photometric measurements were available from AAVSO and ASAS-SN databases and therefore no light curve could be determined.

Its colour index Bp-Rp = 4.993 and absolute magnitude $M_G$ = 6.60 (-1.09, +0.83) are compatible with a YSO. Based on its photometric characteristics, we believe that the original Semiregular type of V1692 Cyg is probably not correct.

The uncertainty of the original period is likely an indicator of an irregular variability typical of a YSO.

**V1693 Cyg**

This is a very red star located 2571 (-855, +1582) pc away with an effective temperature of 3281 (-5, +8) K.

The colour index Bp-Rp = 6.901 and absolute magnitude $M_G$ = 4.00 (-1.04, +0.88) are compatible with a LPV star.

From our analysis of the valid ASAS-SN photometric measurements no reliable solution was found for the period and therefore we cannot provide any indication with respect to the uncertain original period of 400 days.

Our assessment shows that the photometric characteristics are compatible with the original classification, but it is not possible to discriminate between Semiregular or Mira type or refine the period.

**V1694 Cyg**

See assessment of V1638 Cyg.

**V2311 Cyg**

This star is 1249 (-32, +34) pc away, with a temperature of 4897 K (-188, +196), a colour index Bp-Rp = 1.298 and an absolute magnitude $M_G$ = 2.84 ± 0.06. Its photometric and physical data clearly show that this star is not an LPV.

The original classification was already suggesting its binary eclipse system nature, with an amplitude of 0.9 mag in the photographic infrared filter.

The ASAS-SN Catalog of Variable Stars II has improved this classification defining this variable as a β Lyrae system, with a mean magnitude V = 13.09 an amplitude of 0.47 mag. and a period of 1.4114255 days.

Our period analysis with ANOVA method, on 675 valid observations available from ASAS-SN in the V filter and covering a time span of 1358 days, confirmed the β Lyrae type and highlighted a period of 1.4114 ± 0.0001 days (Figure 47), which confirms the ASAS-SN result.

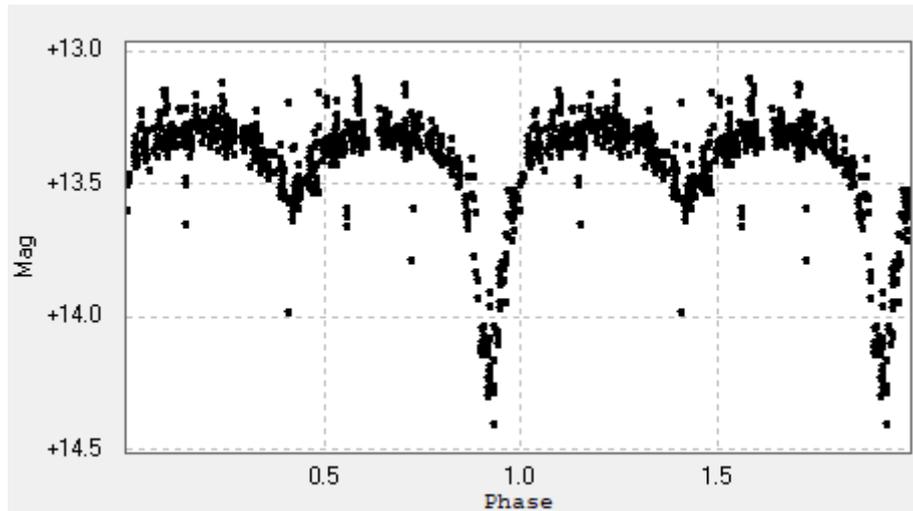

Figure 47      V2311 Cyg – V mag vs Phase (1.4114 ± 0.0001 days period - ANOVA)

## 3. Conclusions

The assessment of data available from public astronomical databases and referenced papers, and the analysis we performed on light curves and periods, refine the original classification, or suggest a revision of the type and/or the period for most of the 62 variables of IBVS 1302.

For all stars, the variability is confirmed.

For the following 17 stars, our analysis provides refinements or solutions for the type and/or period significantly different from existing studies: NSV 25072, V1623 Cyg, V1631 Cyg, V1632 Cyg, V1634 Cyg, V1637 Cyg, V1640 Cyg, V1647 Cyg, V1653 Cyg, V1656 Cyg, V1658 Cyg, V1677 Cyg, V1680 Cyg, V1681 Cyg, V1684 Cyg, V1690 Cyg, V1692 Cyg.

Among other 3 stars, whose original type was unknown and for which no further analysis was available, based on their photometric characteristics, we suggest the membership of NSV 13006 and NSV 25117 to the Semiregular or YSO group and highlight that NSV 25162 is not an LPV Our assessment also identifies 3 cases for which characteristics available from ASAS-SN photometric database are not consistent with results from Gaia DR2 and/or our analysis: V1635 Cyg, classified by ASAS-SN as YSO, unlike the LPV group defined by Gaia DR2 and our analysis, and V1642 Cyg and V1653 Cyg, classified as Semiregular variable with a period of 7 days, which is much shorter than the periods around 300 days found by Gaia DR2 and our analysis.

For variables NSV 25071, V1625 Cyg and V1677 Cyg, our analysis defines the epoch of a maximum for a first time. Overall, our analysis provides the epoch of a minimum or a maximum for 24 variables.

Our assessment also highlights that for 23 stars, incorrect cross-reference names are reported by SIMBAD database and/or ASAS-SN Catalog of Variable Stars with one case of potential misidentification of star V1641 Cyg.

Table 9 provides a summary of the results of this work.

Table 9        Summary of the main results of this work

| | IBVS 1302 | | This assessment | | |
|---|---|---|---|---|---|
| **Variable** | **Type** | **Period (d)** | **Type** | **Period (d)** | **Remarks** |
| NSV 13006 | --- | --- | SR or YSO (our analysis) | --- | Correct cross-reference for ASASSN-V J201904.15+385435.9 is HD 228883 (Gaia DR2 2061308294621232384). |
| NSV 25019 | L:: | --- | LPV (our analysis) | --- | --- |
| NSV 25055 | L:: | --- | LPV (our analysis) <br><br> SR (ASAS-SN) | --- <br><br> 7.6486 (ASAS-SN) | Correct cross-reference for ASASSN-V J201210.10+412224.0 is Gaia DR2 2074619978810093952. |
| NSV 25071 | E: | --- | EA (our analysis) <br><br> EA (ASAS-SN) | 8.8079 ± 0.0027 (our analysis) <br><br> 8.8075552 (ASAS-SN) | Minimum = 2457710.698 ± 0.019 HJD (our analysis) |
| NSV 25072 | C:: | --- | Not LPV (our analysis) | --- | Correct cross-reference for ASASSN-V J201507.25+411751.7 is double star HD 192803. <br> Original uncertain Cepheid type is excluded. |
| NSV 25113 | --- | --- | LPV (our analysis) <br><br> M (AAVSO) | --- <br><br> --- | --- |
| NSV 25117 | --- | --- | SR or YSO (our analysis) | --- | --- |
| NSV 25162 | --- | --- | Not LPV (our analysis) | --- | --- |
| V433 Cyg | M | 400 | LPV (our analysis) <br><br> M (ASAS-SN) <br><br> LPV (Gaia DR2) <br><br> M (AAVSO) | 368 ± 33 (our analysis) <br><br> 369 (ASAS-SN) <br><br> 384 ± 33 (Gaia DR2) <br><br> 417 (AAVSO) | Correct cross-reference for ASASSN-V J201539.86+382545.2 is Gaia DR2 2060941710574890368. <br> Maximum = 2457490 ± 6 HJD (our analysis) |
| V1322 Cyg | L:: | ---- | Not LPV (our analysis) <br><br> GCAS (ASAS-SN & INTEGRAL) | Not applicable | --- |

| | IBVS 1302 | | This assessment | | |
|---|---|---|---|---|---|
| **Variable** | **Type** | **Period (d)** | **Type** | **Period (d)** | **Remarks** |
| V1622 Cyg | M | 350 | LPV (Gaia DR2) | 354 ± 32 (Gaia DR2) | Correct cross-reference for ASASSN-V J200732.14+404345.6 is Gaia DR2 2074170102454813056. Maximum = 2457825 ± 10 (our analysis) |
| | | | LPV (our analysis) | 366 ± 13 (our analysis) | |
| | | | SR (ASAS-SN) | 369 (ASAS-SN) | |
| V1623 Cyg | M | 310 | LPV (our analysis) | 103 ± 3 153 ± 10 222 ± 29 (our analysis) | Correct cross-reference for ASASSN-V J200737.05+395635.3 is Gaia DR2 2074087952635187840. Maximum = 2458384 ± 2 HJD (our analysis) |
| | | | LPV (Gaia DR2) | 306 ± 22 (Gaia DR2) | |
| | | | SR (ASAS-SN) | 310 (ASAS-SN) | |
| V1625 Cyg | SR: | --- | LPV (our analysis) | 464 ± 30 (our analysis) | Correct cross-reference for ASASSN-V J200926.04+404347.2 is Gaia DR2 2074497898646252672. Maximum = 2457669 ± 5 HJD (our analysis) |
| | | | SR (ASAS-SN) | 480 (ASAS-SN) | |
| V1626 Cyg | SR:: | 400 | LPV (our analysis) | Not reliable value (our analysis) | --- |
| | | | SR (ASAS-SN) | 482 (ASAS-SN) | |
| | | | LPV (Gaia DR2) | 626 ± 151 (Gaia DR2) | |
| V1627 Cyg | M | 300 | LPV (our analysis) | 295 ± 11 (our analysis) | Maximum = 2457929 ± 3 HJD (our analysis) |
| | | | SR (ASAS-SN) | 299 (ASAS-SN) | |
| | | | LPV (Gaia DR2) | 310 ± 36 (Gaia DR2) | |
| V1628 Cyg | M | 261 | LPV (our analysis) | 219 ± 37 (our analysis) | Correct cross-reference for ASASSN-V J201051.40+411555.0 is Gaia DR2 2074538958561467008. Maximum = 2457943 ± 2 (our analysis) |
| | | | SR (ASAS-SN) | 257 (ASAS-SN) | |
| | | | LPV (Gaia DR2) | 258 ± 13 (Gaia DR2) | |
| V1629 Cyg | SR | 460: | LPV (our analysis) | 476 ± 67 (our analysis) | Maximum = 2457186 ± 6 HJD (our analysis) |
| | | | SR (ASAS-SN) | 503 (ASAS-SN) | |

| | IBVS 1302 | | This assessment | | |
|---|---|---|---|---|---|
| **Variable** | **Type** | **Period (d)** | **Type** | **Period (d)** | **Remarks** |
| V1630 Cyg | M | 415 | LPV (our analysis) | --- | Correct cross-reference for ASASSN-V J201116.24+410435.0 is Gaia DR2 2074515662659100800. |
| | | | SR (ASAS-SN) | 215 (ASAS-SN) | |
| | | | LPV (Gaia DR2) | 416 ± 76 (Gaia DR2) | |
| V1631 Cyg | M | 430 | LPV (our analysis) | 235 ± 16 (our analysis) | Maximum = 2458030 ± 6 HJD (our analysis) |
| | | | SR (ASAS-SN) | 497 (ASAS-SN) | |
| V1632 Cyg | SR | 318 | SR or YSO (our analysis) | 160 ± 7 425 ± 58 (our analysis) | Correct cross-reference for ASASSN-V J201201.61+410021.3 is Gaia DR2 2074517415005650944. Maximum = 2458250 ± 8 HJD (our analysis) |
| | | | Var (ASAS-SN) | 444 (ASAS-SN) | |
| V1633 Cyg | M | 404 | LPV (Gaia DR2) | 403 ± 49 (Gaia DR2) | Correct cross-reference for ASASSN-V J201208.06+393649.6 is Gaia DR2 2062369288973870976. Maximum = 2458367 ± 4 HJD (our analysis) |
| | | | M (ASAS-SN) | 409 (ASAS-SN) | |
| | | | LPV (our analysis) | 410 ± 8 (our analysis) | |
| V1634 Cyg | M | 251 | LPV (our analysis) | 254 ± 18 (our analysis) | Correct cross-reference for ASASSN-V J201219.38+415149.7 is Gaia DR2 2074717423030592512. Maximum = 2457193 ± 15 HJD (our analysis) |
| | | | YSO (ASAS-SN) | Not applicable (ASAS-SN) | |
| V1635 Cyg | M | 400 | YSO (ASAS-SN) | Not applicable (ASAS-SN) | Most likely correct cross-reference for ASASSN-V J201234.20+390233.8 is Gaia DR2 2061782665176034816, but Gaia DR2 2061782665176035584 cannot be excluded. |
| | | | LPV (our analysis) | --- (our analysis) | |
| | | | LPV (Gaia DR2) | 399 ± 40 (Gaia DR2) | |
| V1636 Cyg | M | 322 | LPV (Gaia DR2) | 297 ± 20 (Gaia DR2) | Correct cross-reference for ASASSN-V J201243.76+413602.8 is Gaia DR2 2074640560281165184. Maximum = 2457487 ± 2 HJD (our analysis) |
| | | | LPV (our analysis) | 320 ± 11 (our analysis) | |
| | | | SR (ASAS-SN) | 326 (ASAS-SN) | |

| | IBVS 1302 | | This assessment | | |
|---|---|---|---|---|---|
| **Variable** | **Type** | **Period (d)** | **Type** | **Period (d)** | **Remarks** |
| V1637 Cyg | SR | 410 | LPV (Gaia DR2) | 212 ± 14 (Gaia DR2) | Correct cross-reference for ASASSN-V J201252.20+410306.7 is Gaia DR2 2062602557236419456. Maximum = 2457660 ± 5 HJD (our analysis) |
| | | | LPV (our analysis) | 267 ± 30 (our analysis) | |
| | | | SR (ASAS-SN) | 424 (ASAS-SN) | |
| V1638 Cyg | SR | 416 | LPV (our analysis) | --- | --- |
| V1639 Cyg | M | 360 | SR (ASAS-SN) | 182 (ASAS-SN) | Correct cross-reference for ASASSN-V J201323.50+383846.7 is Gaia DR2 2061718996577182848. Maximum = 2457929 ± 4 (our analysis) |
| | | | LPV (our analysis) | 182 ± 3 363 ± 10 (our analysis) | |
| | | | LPV (Gaia DR2) | 371 ± 34 (Gaia DR2) | |
| V1640 Cyg | M | 412 | LPV (our analysis) | 29.5 ± 0.13 59.0 ± 0.44 (our analysis) | --- |
| | | | SR (ASAS-SN) | 45.96 (ASAS-SN) | |
| V1641 Cyg | SR or M | 584:: | --- (our analysis) | --- (our analysis) | Potential misidentification: cross-reference for ASASSN-V J201348.91+383807.7 / V1641 to Gaia DR2 2061712193348895872 or 2061712193348896000 may be incorrect. |
| | | | YSO (ASAS-SN) | Not applicable (ASAS-SN) | |
| V1642 Cyg | M | 290 | LPV (our analysis) | --- (our analysis) | --- |
| | | | SR (ASAS-SN) | 7.2068939 (ASAS-SN) | |
| | | | LPV (Gaia DR2) | 296 ± 24 (Gaia DR2) | |
| V1643 Cyg | M | 297 | SR (ASAS-SN) | 289 (ASAS-SN) | Correct cross-reference for ASASSN-V J201420.07+382745.7 is Gaia DR2 2060950884625235584. Maximum = 2457598 ± 4 (our analysis) |
| | | | LPV (Gaia DR2) | 305 ± 23 (Gaia DR2) | |
| | | | LPV (our analysis) | 306 ± 29 (our analysis) | |
| V1645 Cyg | M | 380 | LPV (our analysis) | --- (our analysis) | --- |
| | | | LPV (Gaia DR2) | 354 ± 34 (Gaia DR2) | |

| | IBVS 1302 | | This assessment | | |
|---|---|---|---|---|---|
| **Variable** | **Type** | **Period (d)** | **Type** | **Period (d)** | **Remarks** |
| V1646 Cyg | M: | 388 | --- | --- | No assessment. |
| V1647 Cyg | SR or M | 334: | YSO (our analysis) | Not applicable | --- |
| V1648 Cyg | M | 300 | LPV (our analysis) | No reliable period found (our analysis) | |
| | | | SR (ASAS-SN) | 286 (ASAS-SN) | |
| V1649 Cyg | M | 442 | SR or YSO (our analysis) | No reliable period found (our analysis) | Correct cross-reference for ASASSN-V J201828.99+403406.3 is Gaia DR2 2062342728892073088. |
| | | | SR (ASAS-SN) | 226 (ASAS-SN) | |
| V1650 Cyg | M | 395 | LPV (our analysis) | 385 ± 30 (our analysis) | Correct cross-reference for ASASSN-V J201838.07+391756.0 is Gaia DR2 2061470128971060864. |
| | | | SR (ASAS-SN) | 388 (ASAS-SN) | |
| | | | LPV (Gaia DR2) | 391 ± 58 (Gaia DR2) | |
| V1651 Cyg | M | 383 | LPV (our analysis) | --- | --- |
| V1652 Cyg | M | 565: | SR or YSO (our analysis) | --- | --- |
| V1653 Cyg | M | 335 | SR (ASAS-SN) | 7.06814 (ASAS-SN) | Correct cross-reference for ASASSN-V J202043.11+392239.9 is Gaia DR2 2061416012387664512. Maximum = 2457660 ± 5 (our analysis) |
| | | | LPV (our analysis) | 267 ± 31 (our analysis) | |
| | | | LPV (Gaia DR2) | 314 ± 79 (Gaia DR2) | |
| V1654 Cyg | M | 268: | SR (ASAS-SN) | 252 (ASAS-SN) | Maximum = 2458362 ± 2 HJD (our analysis) |
| | | | LPV (our analysis) | 262 ± 19 (our analysis) | |
| V1655 Cyg | M | 462 | LPV (our analysis) | --- | --- |
| V1656 Cyg | M | 308 | SR (ASAS-SN) | 291 (ASAS-SN) | Maximum = 2457593 ± 7 HJD (our analysis) |
| | | | LPV (our analysis) | 148 ± 9 or 295 ± 28 (our analysis) | |
| | | | LPV (Gaia DR2) | 314 ± 34 (Gaia DR2) | |
| V1657 Cyg | M | 310 | --- | --- | No assessment. |

| | IBVS 1302 | | This assessment | | |
|---|---|---|---|---|---|
| **Variable** | **Type** | **Period (d)** | **Type** | **Period (d)** | **Remarks** |
| V1658 Cyg | M | 382 | LPV (our analysis) | 112 ± 1 (our analysis) | Maximum = 2457895 ± 2 HJD (our analysis) |
| | | | M (ASAS-SN) | 196 (ASAS-SN) | |
| | | | LPV (Gaia DR2) | 413 ± 39 (Gaia DR2) | |
| V1659 Cyg | M | 770 | LPV (our analysis) | No reliable solution (our analysis) | Maximum = 2458904 ± 13 HJD (our analysis) |
| | | | LPV (Gaia DR2) | 491 ± 118 (Gaia DR2) | |
| V1660 Cyg | M | 420 | LPV (our analysis) | --- | --- |
| V1677 Cyg | I | --- | LPV (our analysis) | 29.5 ± 0.2 (our analysis) | Maximum = 2457573 ± 5 HJD (our analysis) |
| | | | Irregular (ASAS-SN) | Not applicable (ASAS-SN) | |
| V1678 Cyg | E | --- | EB (our analysis) | 0.5278 ± 0.0002 (our analysis) | Correct cross-reference for ASASSN-V J201326.99+383432.2 is Gaia DR2 2061706455272623744. |
| | | | EB (ASAS-SN) | 0.52784 (ASAS-SN) | |
| V1680 Cyg | SR | --- | SR or YSO (our analysis) | 215 ± 11 (our analysis) | --- |
| V1681 Cyg | SR | 280: | LPV (our analysis) | 27 ± 0.2 or 55 ± 0.4 (our analysis) | --- |
| V1682 Cyg | SR | 290:: | LPV (our analysis) | --- | --- |
| V1683 Cyg | I | --- | Not LPV (our analysis) | Not reliable period found (our analysis) | --- |
| | | | VAR (ASAS-SN) | --- | |
| V1684 Cyg | SR | 400:: | LPV (our analysis) | 210 ± 14 or 424 ± 25 (our analysis) | Correct cross-reference for ASASSN-V J202003.05+430257.6 is Gaia DR2 2068820948319278976. Maximum = 2458303 ± 10 HJD (our analysis) |
| | | | SR (ASAS-SN) | 395 (ASAS-SN) | |
| | | | LPV (Gaia DR2) | 540 ± 194 (Gaia DR2) | |
| V1688 Cyg | M | 312: | --- | --- | No assessment. |
| V1689 Cyg | SR or M | 484: | --- | --- | No assessment. |
| V1690 Cyg | RV | 285 | LPV (our analysis) | 140 ± 2 or 279 ± 6 (our analysis) | Maximum = 2457962 ± 3 HJD (our analysis) |

| | IBVS 1302 | | This assessment | | |
|---|---|---|---|---|---|
| **Variable** | **Type** | **Period (d)** | **Type** | **Period (d)** | **Remarks** |
| V1691 Cyg | M: | 385: | --- (our analysis) | No reliable period found (our analysis) | --- |
| | | | VAR (ASAS-SN) | --- | |
| V1692 Cyg | SR | 450: | YSO (our analysis) | --- | --- |
| V1693 Cyg | SR or M | 400: | LPV (our analysis) | Not reliable period found (our analysis) | --- |
| V1694 Cyg | SR or M | 308:: | LPV (our analysis) | --- | --- |
| V2311 Cyg | E | --- | EB (our analysis) | 1.4114 ± 0.0001 (our analysis) | --- |
| | | | EB (ASAS-SN) | 1.4114255 (ASAS-SN) | |

**Acknowledgements**
- This activity has made use of the SIMBAD database, operated at CDS, Strasbourg, France.
- This work has made use of data from the European Space Agency (ESA) mission Gaia (https://www.cosmos.esa.int/gaia), processed by the Gaia Data Processing and Analysis Consortium (DPAC, https://www.cosmos.esa.int/web/gaia/dpac/consortium). Funding for the DPAC has been provided by national institutions, in particular the institutions participating in the Gaia Multilateral Agreement.
- We acknowledge with thanks the variable star observations from the *AAVSO International Database* contributed by observers worldwide and used in this research.
- This work was carried out in the context of educational and training activities provided by Italian law 'Percorsi per le Competenze Trasversali e l'Orientamento', December 30<sup>th</sup>, 2018 n.145, Art.1.
**References**
- Alfonso-Garzón et al. 2012, The first INTEGRAL-OMC catalogue of optically variable sources, arXiv:1210.0821v2
- Alksnis et al 2001, General Catalog of galactic Carbon stars, 3d Ed., 2001BaltA..10....1A
- Bailer-Jones et al. 2018, Estimating Distance from Parallaxes. IV. Distances to 1.33 Billion Stars in Gaia Data Release 2, 2018AJ....156...58B
- Cutri et al. 2003, VizieR Online Data Catalog: 2MASS All-Sky Catalog of Point Sources, https://ui.adsabs.harvard.edu/abs/2003yCat.2246....0C/abstract
- Gaia DR2 (Gaia Collaboration, 2018), Gaia Data Release 2. Summary of the contents and survey properties, 2018A&A...616A...1G
- Gaia DR2 (Gaia Collaboration, 2020), Variable stars in the colour-absolute magnitude diagram, https://doi.org/10.1051/0004-6361/201833304
- Gezer et al. 2015, The WISE view of RV Tauri stars, 2015MNRAS.453..133G
- Evans et al. 2018, Gaia Data Release 2 - Photometric content and validation. https://doi.org/10.1051/0004-6361/201832756
- Harris et al. 1989, Photoelectric observations of asteroids 3, 24, 60, 261, and 863. Icarus 77, 171-186
- Horne, J.H., Baliunas, S.L., 1986, A Prescription for Period Analysis of Unevenly Sampled Time Series, 1986ApJ...302..757H
- Jayasinghe et al. 2018b. The ASAS-SN catalogue of variable stars - II. Uniform classification of 412 000 known variables. 2019MNRAS.486.1907J
- Kafka S. 2020, Observations from the AAVSO International Database, https://www.aavso.org/
- Kochanek et al. 2017, The All-Sky Automated Survey for Supernovae (ASAS-SN) Light Curve Server v1.0, arXiv:1706.07060
- Kwok S. et al. (1997), Classification and identification of IRAS sources with low-resolution spectra, 1997ApJS..112..557K
- Laur et al. (2017), Variability survey of brightest stars in selected OB associations, 2017A&A...598A.108L
- Lomb, N.R., 1976, Least-Squares Frequency Analysis of Unequally Spaced Data, 1976Ap&SS..39..447L
- Maffei P. 1977, New Variable Stars in the field of γ Cygni, IBVS 1302, 1977IBVS.1302....1M
58